\documentclass[authoryear,preprint,review,12pt]{elsarticle}

\usepackage{amssymb}
\usepackage{lscape}
\usepackage{caption}
\usepackage{subcaption}
\usepackage{epstopdf}
\usepackage{gensymb}
\usepackage{array}
\usepackage{float}
\usepackage{setspace}
\usepackage{fancyhdr}
\usepackage{amsmath}
\usepackage{longtable}
\usepackage{color}

\journal{Journal of Fluids and Structures}

\begin{document}

\begin{frontmatter}







%\author{}
%\address{}

\title{Wave propagation in stenotic vessels; theoretical analysis and comparison between 3D and 1D fluid-structure-interaction models}

\author[address1]{G. Papadakis \corref{mycorrespondingauthor}}
\address[address1]{Department of Aeronautics, Imperial College London, SW7 2AZ, UK}
\cortext[mycorrespondingauthor]{Corresponding author}
\ead{g.papadakis@ic.c.uk}
\author[address1]{J. Raspaud}

\begin{abstract}
Using analytical expressions for the pressure and velocity waveforms in tapered vessels, we construct a linear 1D model for wave propagation in stenotic vessels in the frequency domain. We demonstrate that using only two parameters to approximate the exact geometry of the constriction (length and degree of stenosis), we can construct a model that can be solved analytically and can approximate with excellent accuracy the  response of the original vessel for a wide range of physiologically relevant frequencies. We then proceed to compare the 1D results with full 3D FSI results from the literature for parameters corresponding to an idealized stenotic carotid artery. We find excellent matching with the volume flow rare over the cardiac cycle (less than $1\%$ error). Using results from DNS simulations to parametrize the velocity profile in the stenotic region, we manage to predict also the pressure distribution with small error (a few percentage points). The method proposed in the paper can be used to approximate vessels of arbitrary shape profile and can be extended to cover the whole cardiovascular tree. Recursive expressions make the solution very fast and open the possibility of carrying out sensitivity and uncertainty quantification studies that require thousands (or even millions) of simulations with minimal cost.
\end{abstract}


\end{frontmatter}


\section{Introduction}\label{sec:introduction}
Stenosis is the narrowing of arteries possibly due to inflammation or plaque buildup (known as atherosclerosis). In the coronary artery, stenosis can cause angina and is an established cause of heart failure; in the renal artery can result in kidney failure. The flow through a constricted vessel is considered as difficult to simulate, mainly because of complexity arising from unsteady flow separation and transition. 

Early research into stenotic flows dates back to 1970's, where experimental in vitro studies for steady and unsteady inflow conditions were conducted in idealized models by \cite{Young_Tsai_1973a,Young_Tsai_1973b}. For a Reynolds number between 100 and 5000 and steady flow conditions the authors encountered three flow regimes: for low Reynolds numbers, the flow was laminar without separation, for larger numbers a zone of back flow was observed, and further increasing the Reynolds number, led to highly turbulent flow and no localized discrete region of reverse flow was discernible. They also found the flow characteristics are strongly affected by the percentage area-reduction of the stenosis. Numerous experimental studies have followed (\cite{Cassanova_Giddens_1978}, \cite{Ahmed_Giddens_1983a}, \cite{ojha_cobbold_johnston_hummel_1989}, \cite{Karri_Vlachos_2010}, \cite{PIELHOP_et_al_2012}, \cite{Choi_et_al_2017} to cite but a few). 

A large number of 2D or 3D numerical simulations have been also conducted. Only 3D simulations can capture the loss of axisymmetry and transition (\cite{sherwin_blackburn_2005}, \cite{varghese_frankel_fischer_2007a,varghese_frankel_fischer_2007b}, \cite{Pal_et_al_2014}). The aforementioned numerical studies ignore the wall distensibility, i.e.\ they assume a rigid wall. Three dimensional models that account for fluid-structure-interaction are those of \cite{Gerbeau_et_al_2005} and \cite{Figueroa_et_al_2006}. Fully resolved 3D simulations are not suitable for modelling the whole arterial network. For this purpose, simpler 1D models have been successfully employed (\cite{Mynard_Nithiarasu_2008}, \cite{Alastruey_et_al_2011}, \cite{Sherwin_et_al_2003}). These models are formulated in physical, i.e.\ space-time, variables. More details on modeling studies (as well as experiments) on steady and unsteady flows in arteries are provided in the review articles of \cite{Berger_Jou_2000}, \cite{Ku_1997}, \cite{Taylor_Draney_2004} and \cite{van_de_Vosse_Stergiopoulos_2011}.

The central aim of the paper is to derive, apply and validate a linear 1D model for a stenotic vessel in terms of space-frequency (as opposed to space-time) variables. The model is based on analytical solutions of pressure and velocity waveforms in elastic, tapered vessels that were derived by \cite{Papadakis_2011}. These solutions were found to match very well with 2D fluid-structure interaction (FSI) results. We present here the application of these solutions to model wave propagation in stenotic vessels and we validate the results against 3D FSI solutions. 

The most significant advantage of the proposed analytical, frequency domain, solution method is that it is very fast. When the equations are formulated in space-time variables, the large speed of blood pulse waves and short lengths of vessels, restricts the time step to very small values, to the order of $10^{-4}s$ (\cite{Du_et_al_2016}) or even less. Furthermore, one needs to evolve the solution for a sufficient number of cardiac cycles (each with period around $0.8s-1s$) to obtain an asymptotic periodic result. The combination of small time step and long integration times leads to an appreciable overall computing time. 

For some applications, many simulations of the 1D model (thousands or even millions) may be necessary. Uncertainty quantification (see \cite{Chen_et_al_2013}) and sensitivity analysis due to lack of accurate information on vessel properties, both require the application of the 1D model multiple times. To give an example, \cite{Huberts_al_2013} investigated the effect of uncertainty of 73 parameters of the pulse propagation model. Their objective was to determine the most influential of these parameters when the model is applied to support decision making in arteriovenous fistula (AVF) surgery for hemodialysis. Their computational framework involved a variance-based method and Monte Carlo simulations. In total, they had to perform more than 2 million simulations, each taking an average 20s. 

This is exactly where frequency domain methods offer a distinct advantage. For example, \cite{Flores_et_al_2016} presented and validated a linear 1D formulation of blood flow in networks of flexible vessels for which a full analytical solution exists in the frequency domain. They call this the generalized Darcy's elastic model (GDEM). The model is based on a linear relationship of the volume flow rate and pressure gradient in the frequency domain, known as Darcy's law. Using appropriate compatibility conditions between successive vessels and at bifurcation points, they derive a linear system of equations with unknowns the values of pressure at the nodes of the cardiovascular tree (a node is defined as the mathematical point where two or three vessels are connected). Our approach is along the lines of \cite{Flores_et_al_2016}, and can be extended to cover the whole cardiovascular tree. 

The paper is organized as follows. In section \ref{sec:1Dmodel}, we explain how the  analytical solutions derived for tapered vessels can be used to model wave propagation in stenotic vessels, and in section \ref{sec: freq_response} we apply the method to compute the frequency response of a vessel. In section \ref{sec:theoretical_analysis}, we simplify the form of the stenotic region and show that we can derive analytically solutions that match very closely with the results of the full model for physiologically relevant frequencies (section \ref{sec:comparison_analytical_numerical}). In section \ref{sec:comparison_3D_1D}, we compare the results of the developed 1D model against benchmark 3D solutions, and we conclude in section \ref{sec:conclusion}.

\section{Mathematical formulation of 1D wave propagation model in a stenotic vessel}\label{sec:1Dmodel}
In this section we propose an analytical 1D model for wave propagation in a stenotic vessel. The profile of the vessel is approximated by a series of $N$ tapered segments (or elements), each of length $\Delta x$, as shown in figure \ref{fig:segmentation} below. The $i$-th element is defined by the boundary faces $i-1$ and $i$, $i =1...N$, refer to figure \ref{fig:ith-element}. 

\begin{figure}[!htb]
\centering
\begin{subfigure}[b]{0.99\textwidth}
\includegraphics[width=0.99\textwidth]{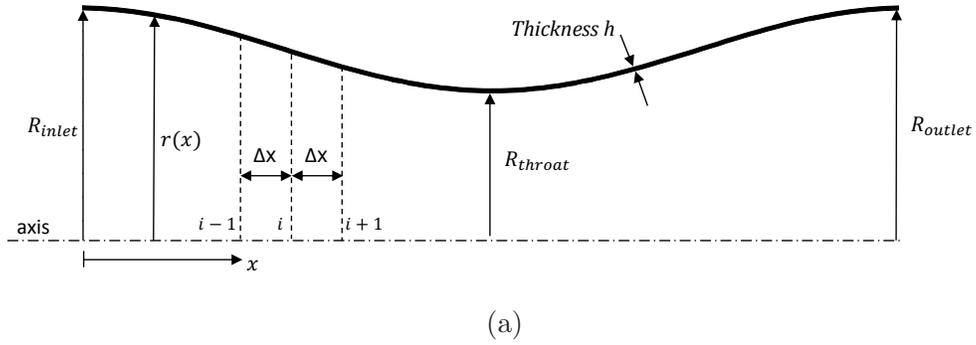}
\caption{}
\label{fig:segmentation}
\end{subfigure}
~
\begin{subfigure}[b]{0.99\textwidth}
\includegraphics[width=0.99\textwidth]{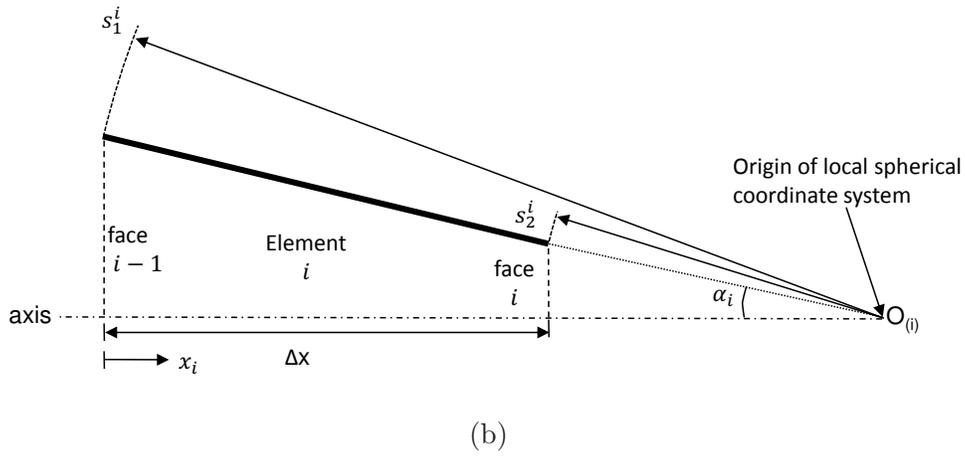}
\caption{}
\label{fig:ith-element}
\end{subfigure}
\caption{(a) Representation of a stenotic vessel as a union of tapering segments (elements), (b) Geometrical details of the $i$-th element.}
\end{figure}

The governing equations for each element take the form (for details refer to \cite{Papadakis_2011}):
\begin{subequations}
\begin{align}
& \frac{\partial}{\partial s}\left(s^2u_s\right)+\left(fs^3\right)\frac{\partial p}{\partial t} = 0 \label{eq:continuity_sph} \\
& \frac{\partial u_s}{\partial t} =  -\frac{1}{\rho}\frac{\partial p}{\partial s} 
\label{eq:momentum_sph}
\end{align}
\label{eq:continuity_momentum_sph}
\end{subequations}
\noindent The equations are formulated in a local spherical reference frame, and the radius $s$ is measured from the origin of the coordinate system $O_{(i)}$, as shown in figure \ref{fig:ith-element}. For the converging element $i$ shown in the figure, $s_2^i\le s \le s_1^i$, where the subscripts $1$ and $2$ refer to flow inlet and outlet respectively. Velocity $u_s$ (positive in the $s$ direction) and pressure $p$ are cross-section averaged and $\rho$ is the density of the fluid. The area of the cross section is $A(s)=2\pi (1-\cos\alpha) s^2$ where $\alpha$ is the (local) cone angle. The flow is along the positive $x$ direction, so in the converging part of the vessel, $s$ and $x$ are in opposite directions, while in the expanding part they are at the same direction. Parameter $f$ in equation \eqref{eq:continuity_sph} is a constant, and is defined as: 
\begin{equation}
f=\frac{(1-\nu^2)\tan^2{\alpha}\sin{\alpha}}{Eh(1-\cos{\alpha})}
\label{eq:f_parameter}
\end{equation}
\noindent It groups together the geometrical and wall parameters of the local segment (the tapering angle $\alpha$, the modulus of elasticity $E$, the thickness of the vessel wall $h$, and the Poisson ratio $\nu$).

Introducing the volume flux $Q_s=A(s)u_s=2\pi (1-\cos\alpha) s^2 u_s$, system \eqref{eq:continuity_momentum_sph} can be written in matrix form as
\begin{equation}
\frac{\partial}{\partial t} 
\begin{bmatrix}
p \\
Q_s 
\end{bmatrix} 
+
\begin{bmatrix}
0                                    & \frac{1}{2\pi(1-\cos\alpha)f s^3} \\
2\pi(1-\cos\alpha) \frac{s^2}{\rho} & 0
\end{bmatrix}
\frac{\partial}{\partial s} 
\begin{bmatrix}
p \\
Q_s 
\end{bmatrix} 
=
\begin{bmatrix}
0 \\
0 
\end{bmatrix} 
\label{eq:continuity_momentum_matrix}
\end{equation}
The eigenvalues of the coefficient matrix are the propagation velocities $\lambda_{1,2}=\pm \left( fs \rho \right)^{-1/2}$. It can be easily shown that the matrix of eigenvectors $R(s)$ can be written as 
\begin{equation}
R(s)=
\begin{bmatrix}
-Z(s) & Z(s) \\
1     & 1    
\end{bmatrix} 
\end{equation}
\noindent where $Z(s)=\dfrac{1}{2\pi(1-\cos\alpha)} \sqrt{\dfrac{\rho}{f}} s^{-5/2} $ is called resistance. The reciprocal  
\begin{equation}
Y(s)=\frac{1}{Z(s)}=2\pi(1-\cos\alpha) \sqrt{\frac{f}{\rho}} s^{+5/2} 
\label{eq:admittance}
\end{equation}
\noindent is called admittance. The characteristic variables $w_1,w_2$ propagating in the positive and negative $s$ directions respectively are,
\begin{equation}
w_1(s)=\frac{1}{2\lambda \rho} \left( p+\lambda \rho u_s\right), \quad w_2(s)=\frac{1}{2\lambda \rho} \left(-p+\lambda \rho u_s\right)
\label{eq:w1_w2}
\end{equation}
These variables are not invariants along a characteristic as in straight vessels; they are instead fully coupled because of the continuous reflection from the tapering wall.

It was shown in \cite{Papadakis_2011} that if the system of equations \eqref{eq:continuity_momentum_sph} is formulated in the frequency domain, i.e. we assume a solution of the form, 
\begin{equation}
p\left(x,t\right)=P\left(x,\omega\right) e^{i\omega t}, \ \ u_s\left(x,t\right)=U_s\left(x,\omega\right)e^{i\omega t},
\label{eq:p_u_freq_domain}
\end{equation}
\noindent then closed form analytic expressions can be obtained for the complex amplitudes $P\left(x,\omega\right)$ and $U_s\left(x,\omega\right)$, as follows
\begin{subequations}
\begin{align}
P\left(x,\omega\right) & =\frac{1}{\sqrt{s}}\left(A{\ J}_\frac{1}{3}\left(z\right)+B{\ Y}_\frac{1}{3}\left(z\right)\right) \label{eq:analytical_solution_p} \\
U_s\left(x,\omega\right) & ={\frac{1}{i}\left(\frac{f}{\rho}\right)}^\frac{1}{2}\left(A\ J_\frac{4}{3}\left(z\right)+B{\ Y}_\frac{4}{3}\left(z\right)\right)
\end{align}
\label{eq:analytical_solutions}
\end{subequations}
\noindent and after some algebra
\begin{equation}
Q_s\left(x,\omega\right)=-iY(s)\frac{1}{\sqrt{s}}\left(A\ J_\frac{4}{3}\left(z\right)+B{\ Y}_\frac{4}{3}\left(z\right)\right)
\label{eq:analytical_solution_Qs}
\end{equation}

\noindent In the above expressions, ${\ J}_\frac{1}{3}\left(z\right), {\ Y}_\frac{4}{3}\left(z\right)$ are Bessel functions of order $\frac{1}{3}$ and $\frac{4}{3}$ respectively, with real argument $z=\frac{2}{3}\omega{(\rho f)}^\frac{1}{2}s^\frac{3}{2}$. Constants $A$ and $B$ are determined by the boundary conditions imposed at the two ends of each segment, locations $s=s_1^i$ and $s=s_2^i$ (refer to figure \ref{fig:ith-element}). These expressions incorporate analytically the change in amplitude of the forward and backward reflecting waves due to tapering. In the work of \cite{Flores_et_al_2016}, a linear relation between the volume flow rate and pressure gradient is applied (Darcy's law). Such a relation is valid for cylindrical vessels only; for tapered vessels such an expression cannot be derived. 

In total there are 2N unknown constants, the set of coefficients $A$ and $B$ for each segment (denoted below as $A_i$ and $B_i$ respectively for the $i-th$ segment). To evaluate these coefficients we apply 
\begin{itemize}
\item $2(N-1)$ compatibility conditions for velocity and pressure at the interfaces between two consecutive segments, and
\item $2$ boundary conditions at the inlet and outlet of the whole vessel.
\end{itemize}

Figure \ref{fig:same_direction} below shows two consecutive elements, $i$ and $i+1$, that share the $i$-th face and taper in the same direction (they both converge for increasing $x$). If the common face is located at the throat of the vessel, then the elements taper in opposite directions (i.e. one will converge and the other will expand), as shown in figure \ref{fig:opposing_direction}.
\begin{figure}[!htb]
\centering
\begin{subfigure}[b]{0.99\textwidth}
\includegraphics[width=0.99\textwidth]{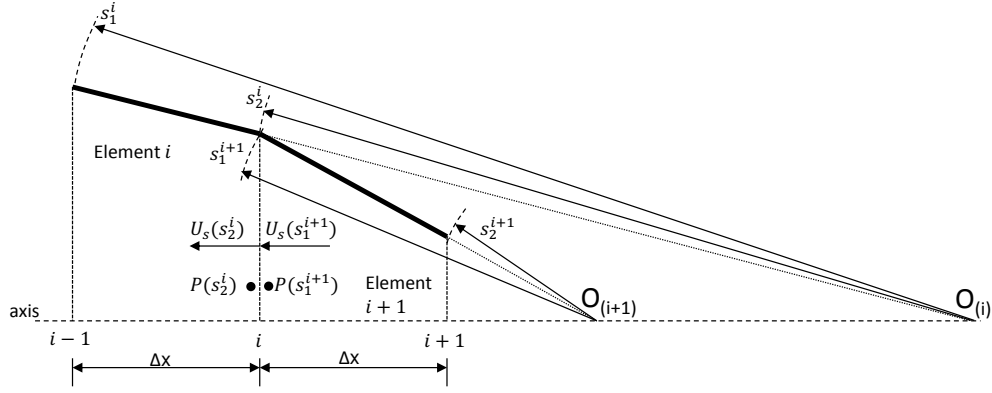}
\caption{}
\label{fig:same_direction}
\end{subfigure}
~
\begin{subfigure}[b]{0.99\textwidth}
\includegraphics[width=0.99\textwidth]{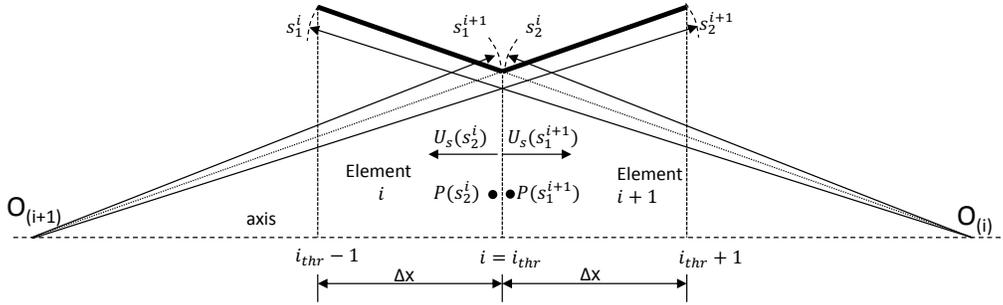}
\caption{}
\label{fig:opposing_direction}
\end{subfigure}
\caption{Two consecutive segments that taper in (a) the same direction and (b) opposing directions (throat location).}
\label{fig:consecutive_segments}
\end{figure}
	
As mentioned earlier, radii $s_1^i$ and $s_2^i$ denote the inlet (upstream) and outlet (downstream) edges of the segment $i$ respectively. The compatibility condition for pressure between segments $i$ and $i+1$ leads to 
\begin{equation}
\begin{gathered}
 P(s_2^i)=P(s_1^{i+1}) \Rightarrow \\
 \frac{1}{\sqrt{s_1^i}}\left(A_iJ_\frac{1}{3}\left(z_2^i\right)+B_iY_\frac{1}{3}\left(z_2^i\right)\right)=\frac{1}{\sqrt{s_1^{i+1}}}\left(A_{i+1}J_\frac{1}{3}\left(z_1^{i+1}\right)+B_{i+1}Y_\frac{1}{3}\left(z_1^{i+1}\right)\right)
\end{gathered}
\label{eq:p_compatibility}
\end{equation}

Similarly for velocity we get the condition
\begin{equation}
\begin{gathered}
 U_s(s_2^i)=U_s(s_1^{i+1}) \Rightarrow \\
 {\left(\frac{f^{i}}{\rho}\right)}^\frac{1}{2}\left(A_{i}J_\frac{4}{3}\left(z_2^{i}\right)+B_{i}Y_\frac{4}{3}\left(z_2^{i}\right)\right)=
{\left(\frac{f^{i+1}}{\rho}\right)}^\frac{1}{2}\left(A_{i+1} J_\frac{4}{3}\left(z_1^{i+1}\right)+B_{i+1} Y_\frac{4}{3}\left(z_1^{i+1}\right)\right)
\end{gathered}
\label{eq:u_compatibility}
\end{equation}

For the interface located exactly at the throat, i.e.\ for $i=i_{thr}$ (refer to figure \ref{fig:opposing_direction}), a change of sign is necessary for velocity i.e.  
\begin{equation}
\begin{gathered}
U_s(s_2^{i_{thr}})=-U_s(s_1^{i_{thr}+1}) \Rightarrow \\
\left(\frac{f^{i_{thr}}}{\rho}\right)^\frac{1}{2} \left( A_{i_{thr}} 
J_\frac{4}{3} \left(z_2^{i_{thr}}\right)+ B_{i_{thr}} Y_\frac{4}{3}\left(z_2^{i_{thr}}\right) \right)=  \\
-\left(\frac{f^{i_{thr}+1}}{\rho}\right)^\frac{1}{2} \left(A_{i_{thr}+1} J_\frac{4}{3}\left(z_1^{i_{thr}+1}\right)+B_{i_{thr}+1} Y_\frac{4}{3}\left(z_1^{i_{thr}+1}\right)\right)
\end{gathered}
\label{eq:u_compatibility_throat}
\end{equation}

At the inlet we assume that pressure is prescribed i.e.\
\begin{equation}
\frac{1}{\sqrt{s_1^1}}\left(A_1J_\frac{1}{3}\left(z_1^1\right)+B_1Y_\frac{1}{3}\left(z_1^1\right)\right)=P_0
\end{equation}
\noindent where $P_0=1$, i.e.\ unit pressure excitation. For the outlet boundary, we assume that there is no incoming wave (non-reflecting condition) i.e.\ the characteristic variable $w_2(s_1^N)=0$ in \eqref{eq:w1_w2}, which leads to 
\begin{equation}
\begin{gathered}
w_2(s_1^N)=0 \Rightarrow -P \left( s_1^N \right)+\lambda \left( s_1^N \right) \rho U_s \left(s_1^N \right) =0  \Rightarrow \\
-\frac{1}{\sqrt{s_1^N}}\left(A_NJ_\frac{1}{3}\left(z_1^N\right)+B_NY_\frac{1}{3}\left(z_1^N\right)\right)+ \\
\left(\rho f^Ns_1^N\right)^{-\frac{1}{2}} \rho \left[ \frac{1}{i} \left(\frac{f^N}{\rho}\right)^\frac{1}{2}\left(A_NJ_\frac{4}{3}\left(z_1^N\right)+B_NY_\frac{4}{3}\left(z_1^N\right)\right) \right]=0
\end{gathered}
\end{equation}

Other boundary conditions can be easily implemented at the inlet and outlet. For example, in section \ref{sec:comparison_3D_1D} we apply a prescribed velocity at the inlet and impedance boundary condition at the outlet. The resulting system of equations for $A_i,B_i$ has a lower block diagonal structure, with block size $2\times2$. Although in practice it was solved numerically, the structure of the system allows a recursive computation, with successive elimination of unknowns starting from the $N-th$ segment and proceeding backwards to the first. This is an important observation and will be evoked later in section \ref{sec:theoretical_analysis} where a closed form analytical expression is derived for a simplified stenotic vessel profile. We note that in the work of \cite{Flores_et_al_2016}, the linear system does not have this convenient matrix structure, and can be solved only numerically. 

\section{Frequency response of a stenotic vessel}\label{sec: freq_response}
\begin{figure}[!htb]
\includegraphics[width=0.99\textwidth]{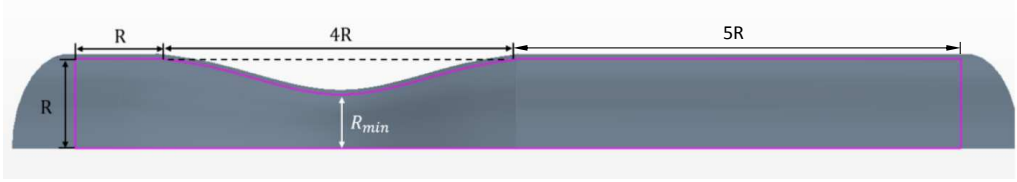}
\caption{Geometric sketch of stenotic vessel}
\label{fig:vessel_sketch}
\end{figure}

We apply the formulation of the previous section to compute the frequency response of a stenotic vessel that consists of three parts: a straight duct of length and radius $R=1cm$, a symmetric converging-diverging nozzle with a total length $4R$ and a straight duct with length $5R$. A sketch is shown in figure \ref{fig:vessel_sketch} and table \ref{tab: vessel_parameters} summarises the fluid and wall properties. These properties are valid for human vessels such as the splenic veins, \cite{Unsal_et_al_2006}.
 
\begin{table}[!htb]
\centering
\begin{tabular}{|l|l|}
\hline
 Fluid Density $\rho_f (kg/m^3)$ & $1000$ \\
\hline
 Wall Density $\rho_s (kg/m^3)$ & $1000$ \\
\hline
 Modulus of elasticity $E (Pa)$ & $10^6$ \\ 
\hline
Poisson ratio $\nu (-)$ & 0.3 \\
\hline
Wall Thickness $h (m)$  &  $0.5 \times 10^{-3}$ \\
\hline
\end{tabular}
\caption{Fluid and wall properties.}
\label{tab: vessel_parameters}
\end{table}

The vessel is discretised using $N=100$ tapered segments, each $1mm$ long. The converging-diverging part is approximated with $40$ elements (these values lead to grid independent results). The two straight parts are approximated as tapered vessels with a very shallow angle ($\alpha=0.001^\circ$) to avoid a singularity in the evaluation of parameter $f$ (refer to \eqref{eq:f_parameter}). The linear system was formulated and solved with $LU$ decomposition in MATLAB.

The degree of stenosis, $S$, is defined as the normalized cross-sectional area reduction between the throat and the inlet i.e.\ $S=1-\left(\frac{R_{min}}{R}\right)^2$. Three cases are considered with $S$ equal to $36\%$, $51\%$ and $64\%$. For each case, the vessel profile was obtained by fitting a $6th$ order polynomial through  the following 7 points with (axial, radial) coordinates equal to $(0,R)$, $(R,R)$, $(2R,R_{mid})$, $(3R,R_{min})$, $(4R,R_{mid})$, $(5R,R)$ and $(6R,R)$, where $R_{mid}=0.5(R+R_{min})$. A straight part with shallow angle was then added to extend the vessel to a total length of $10R$. 

\begin{figure}[!htb]
\includegraphics[width=\textwidth]{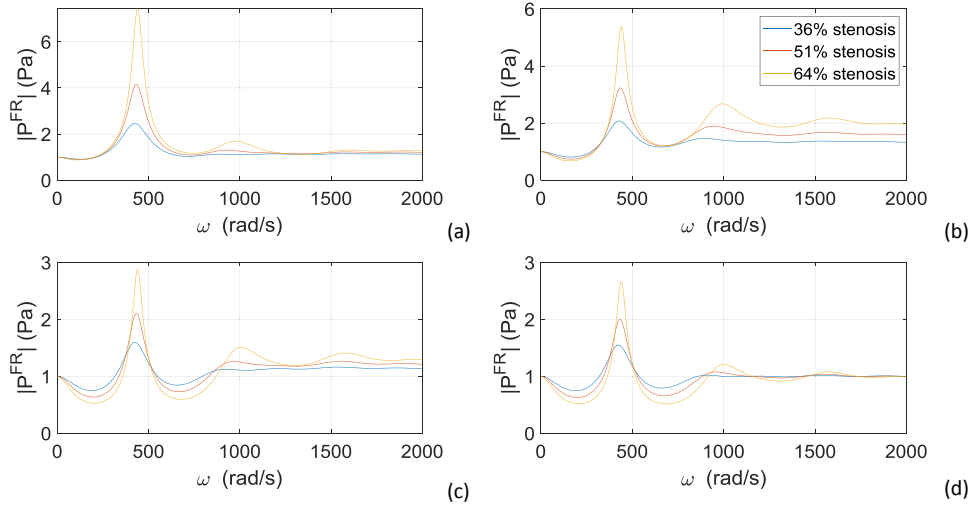}
\caption{Magnitude of pressure response against angular frequency at (a) middle of the converging part of the stenotic region ($x=20mm$), (b) the throat ($x=30mm$), (c) the middle of the diverging part ($x=40mm$) and (d) the end of the stenotic part ($x=50mm$).}
\label{fig:pressure_kernel_3stenoses}
\end{figure}

\begin{figure}[!htb]
\includegraphics[width=\textwidth]{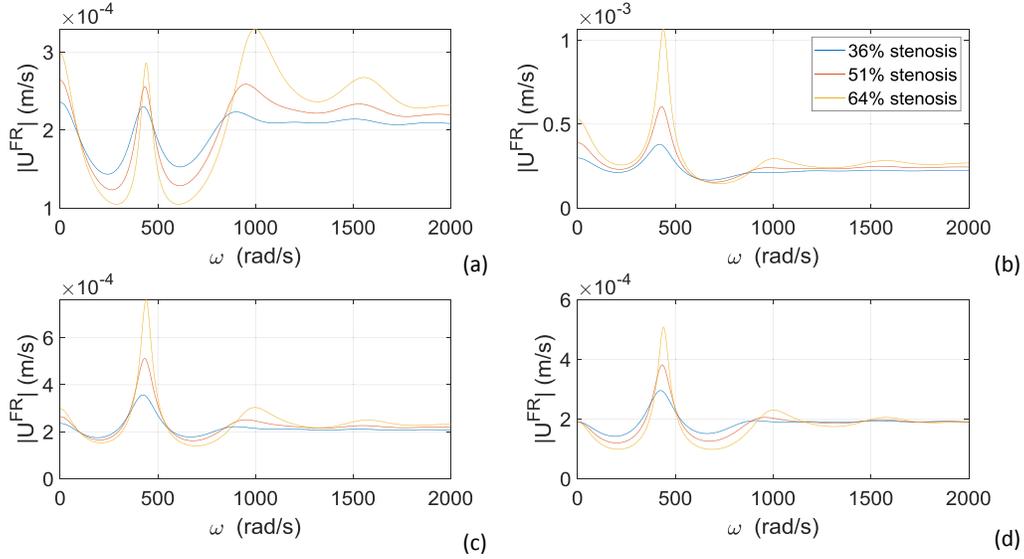}
\caption{Magnitude of velocity response against angular frequency at (a) middle of the converging part of the stenotic region ($x=20mm$), (b) the throat ($x=30mm$), (c) the middle of the diverging part ($x=40mm$) and (d) the end of the stenotic part ($x=50mm$).}
\label{fig:velocity_kernel_3stenoses}
\end{figure}

The magnitude of the pressure and velocity response is plotted against the angular frequency $\omega$ at 4 spatial locations in figures \ref{fig:pressure_kernel_3stenoses} and \ref{fig:velocity_kernel_3stenoses} respectively. The superscript “$FR$” in $P^{FR}$ and $U^{FR}$ in the vertical axes denotes ''Frequency Response'' for unit pressure excitation at the inlet, i.e.\ $P_0=1Pa$.

In order to interpret the results, we consider the angular frequency corresponding to a wavelength equal to a characteristic length scale of the vessel. For the case considered, we take this length scale to be $5R$; this is a dynamically important scale because there is no reflection in the straight part between $5R-10R$. The wave propagation velocities at the straight parts and the throat are equal to $c_{str}=\sqrt{\frac{Eh}{(1-\nu^2)\rho(2R)}}=5.24m/s$ and $c_{thr}=\sqrt{\frac{Eh}{(1-\nu^2)\rho(2R_{min})}}=5.86m/s, 6.26m/s$ and $6.76m/s$ for $S=36\%, 51\%$ and $64\%$ respectively. Considering the average velocities $c_{ave}=5.55 m/s, 5.75m/s$ and $6.0m/s$ for the 3 values of $S$ and length of $5R$ we get the reference angular frequencies $\omega_{ref}=\frac{2\pi c_{ave}}{5R}=697 rad/s,722 rad/s$ and $753 rad/s$. 

Both pressure and velocity plots show resonance peaks in a narrow frequency range, $[436, 442] rad/s$. In this range, the wavelength of the inlet excitation is of the same order as the length scale of the vessel and resonance occurs. This is due to the superposition of the forward and backward moving waves from the converging and diverging parts. Backward waves also reflect at the inlet and propagate forward. At every location therefore there is a multitude of waves that are superimposed. If they are in phase, a constructive interference occurs resulting in a significant amplification of the pressure, i.e.\ resonance. As expected, the amplification increases with the stenosis degree $S$. It also depends on the location along the vessel;  the converging part of the vessel amplifies the pressure wave more compared to the diverging part. Indeed, the maximum amplification appears at the middle of the converging part and it is equal to 7.5 for $S=64\%$. Regarding velocity, the maximum amplification is located at the throat, as it can be expected (refer to figure \ref{fig:velocity_kernel_3stenoses}). 

The frequency range at which resonance occurs is however much larger that the physiological range, say between $[0,50] rad/s$. For such low frequencies, the wavelength is much larger that the characteristic length scale of the vessel and the interaction is much weaker.   

In order to obtain a deeper physical insight on the interaction between the forward and reflected waves in a stenotic vessel, in the following section we derive an analytical solution by simplifying the profile shape. We then compare the results of the simplified model to the full model in section \ref{sec:comparison_analytical_numerical}.  

\section{Analytical solution for a simplified stenotic vessel.}
\label{sec:theoretical_analysis}
We consider the simplified profile shape shown in figure \ref{fig:simplified_stenotic_vessel} below. We have retained only the two most important geometric features, the length of the stenotic section and the degree of stenosis, $S$ (marked with solid blue line in the figure). We replace the detailed profile with two tapered and two straight segments, and therefore the whole vessel consists of 4 compartments. As will be shown below, this simplification allows us to derive a recursive analytic solution by combining expressions valid for straight and tapered sections. For the case examined in the previous section, the characteristic interface locations $x_0,x_1,x_2,x_3,x_4$ take the values $0,R,3R,5R$ and $10R$ respectively. 

\begin{figure}[!htb]
\includegraphics[width=\textwidth]{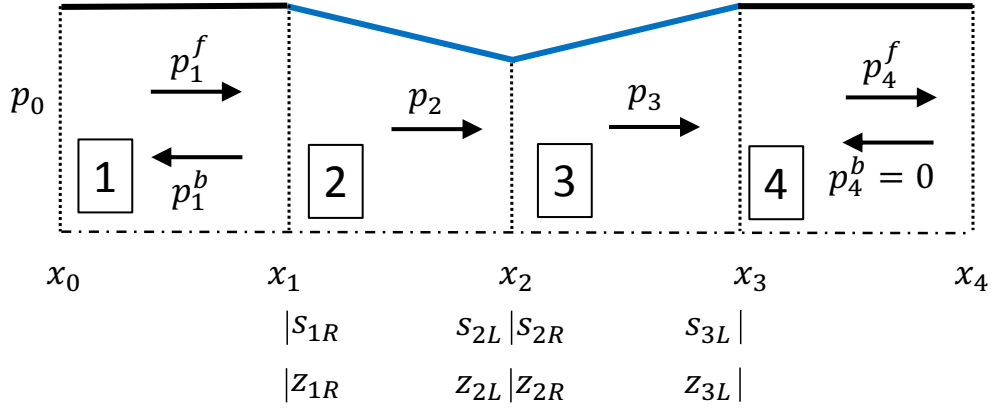}
\caption{Sketch of a simplified stenotic vessel consisting of 4 compartments.}
\label{fig:simplified_stenotic_vessel}
\end{figure}

For the analytic solution, we need also the spherical radii of the inlet and outlet of each tapered compartment. These are denoted by ``$s$'' followed by the index of the interface location and the letter ``L'' or ``R'', depending on whether they were computed from the left or the right tapered section, see figure \ref{fig:simplified_stenotic_vessel}. For example, $s_{2L}$ is the spherical radius evaluated at interface 2 (between compartments 2 and 3) and is referring to the outlet of the left compartment (i.e. 2). The corresponding argument of the Bessel function is denoted as $z_{2L}=\frac{2}{3}\omega{(\rho f_2)}^\frac{1}{2}s_{2L}^{\frac{3}{2}}$. 

Starting with compartment 4, we can write (see \cite{Lighthill_1975}):
\begin{subequations}
\begin{align}
& p_4\left(x,t\right)=p_4^fe^{i\omega(t-\frac{x-x_3}{c_4})}+p_4^be^{i\omega(t+\frac{x-x_3}{c_4})}=p_4^fe^{i\omega(t-\frac{x-x_3}{c_4})} \\
& Q_4\left(x,t\right)=Y_4\left(p_4^fe^{i\omega(t-\frac{x-x_3}{c_4})}-p_4^be^{i\omega(t+\frac{x-x_3}{c_4})}\right)=Y_4p_4^fe^{i\omega(t-\frac{x-x_3}{c_4})}\ 
\end{align}
\end{subequations}

\noindent where $p_4^f$ and $p_4^b$ are the (constant) amplitudes of the forward and backward propagating waves, $c_4$ is the wave propagation velocity and $Y_4=\frac{A_4}{\rho c_4}$ is the admittance of compartment 4. Due to the non-relective boundary condition at $x=x_4$, amplitude $p_4^b=0$.  

For the tapering compartment 3, we apply the analytic solutions \eqref{eq:analytical_solution_p} and \eqref{eq:analytical_solution_Qs},
\begin{subequations}
\begin{align}
& p_3\left(s,t\right)=s^{-\frac{1}{2}}\left(A_3{\ J}_\frac{1}{3}\left(z\right)+B_3{\ Y}_\frac{1}{3}\left(z\right)\right)e^{i\omega t} \\
& {Q_s}_3\left(s,t\right)=-iY(s)s^{-\frac{1}{2}}\left(A_3\ J_\frac{4}{3}\left(z\right)+B_3{\ Y}_\frac{4}{3}\left(z\right)\right)e^{i\omega t}
\end{align}
\end{subequations}

Applying continuity of flow rate and pressure at the interface $x_3$ between compartments 3 and 4 we get
\begin{subequations}
\begin{align}
& p_4^f=s_{3L}^{-\frac{1}{2}}\left(A_3{\ J}_\frac{1}{3}\left(z_{3L}\right)+B_3{\ Y}_\frac{1}{3}\left(z_{3L}\right)\right) \\
& Y_3p_4^f=-iY(s_{3L})s_3^{-\frac{1}{2}}\left(A_3\ J_\frac{4}{3}\left(z_{3L}\right)+B_3{\ Y}_\frac{4}{3}\left(z_{3L}\right)\right)
\end{align}
\end{subequations}

Assuming that the admittance at the interface $x_3$ is continuous, i.e.\ $Y_3=Y(s_{3L})$, and eliminating $p_4^f$ we get
\begin{equation}
\frac{B_3}{A_3}=-\frac{iJ_\frac{4}{3}\left(z_{3L}\right)+{\ J}_\frac{1}{3}\left(z_{3L}\right)}{{\ Y}_\frac{1}{3}\left(z_{3L}\right)+i{\ Y}_\frac{4}{3}\left(z_{3L}\right)}
\label{eq:B3_over_A3}
\end{equation}

\noindent Strictly speaking, we expect a weak discontinuity of $Y$ at $x_3$ because of the slope change. We chose to ignore it here in order to simplify the resulting analytical expressions.

The effective admittance at $x_2$ is defined as $Y_{eff}\left(x_2\right)=\dfrac{Q_{s3}(s_{2R},t)}{p_3(s_{2R},t)}$ and is equal to 
\begin{equation}
Y_{eff}\left(x_2 \right)=-iY\left(s_{2R}\right)\frac{\ J_\frac{4}{3}\left(z_{2R}\right)+\frac{B_3}{A_3}{\ Y}_\frac{4}{3}\left(z_{2R}\right)}{{\ J}_\frac{1}{3}\left(z_{2R}\right)+\frac{B_3}{A_3}{\ Y}_\frac{1}{3}\left(z_{2R}\right)}
\label{eq:Yeffx2r}
\end{equation}

\noindent with the ratio $\frac{B_3}{A_3}$ given by \eqref{eq:B3_over_A3}. We proceed in the same way for compartment 2, where
\begin{subequations}
\begin{align}
& p_2\left(s,t\right)=s^{-\frac{1}{2}}\left(A_2{\ J}_\frac{1}{3}\left(z\right)+B_2{\ Y}_\frac{1}{3}\left(z\right)\right)e^{i\omega t} \\
& {Q_s}_2\left(s,t\right)=-iY(s)s^{-\frac{1}{2}}\left(A_2\ J_\frac{4}{3}\left(z\right)+B_2{\ Y}_\frac{4}{3}\left(z\right)\right)e^{i\omega t}
\end{align}
\end{subequations}

and apply continuity at the interface $x_2$
\begin{subequations}
\begin{align}
& s_{2L}^{-\frac{1}{2}}\left(A_2{\ J}_\frac{1}{3}\left(z_{2L}\right)+B_2{\ Y}_\frac{1}{3}\left(z_{2L}\right)\right)e^{i\omega t}=p_3(s_{2R},t) \\
& iY\left(s_{2L}\right)s_{2L}^{-\frac{1}{2}}\left(A_2\ J_\frac{4}{3}\left(z_{2L}\right)+B_2{\ Y}_\frac{4}{3}\left(z_{2L}\right)\right)={Q_s}_3(s_{2R},t)=Y_{eff}\left(x_2\right)p_3(s_{2R},t) \label{eq:Q_compatibility_simple}
\end{align}
\end{subequations}

Notice that in the left hand side of \eqref{eq:Q_compatibility_simple} the sign was changed from $(-)$ to $(+)$. This is because $x_2$ is located at the throat of the stenotic vessel and the spherical radii in compartments 2 and 3 grow in opposite directions (this equation is the exact analogue of \eqref{eq:u_compatibility_throat}). We take as positive the +x direction.  After some algebra we get
\begin{equation}
\frac{B_2}{A_2}=\frac{iJ_\frac{4}{3}\left(z_{2L}\right)-\frac{Y_{eff}\left(x_{2}\right)}{Y\left(s_{2L}\right)}{\ J}_\frac{1}{3}\left(z_{2L}\right)}{\frac{Y_{eff}\left(x_{2R}\right)}{Y\left(s_{2L}\right)}{\ Y}_\frac{1}{3}\left(z_{2L}\right)-i{\ Y}_\frac{4}{3}\left(z_{2L}\right)}
\label{eq:B2_over_A2}
\end{equation}

\noindent where the ratio $\frac{Y_{eff}\left(x_{2R}\right)}{Y\left(s_{2L}\right)}$ can be computed from equation \eqref{eq:Yeffx2r} above, because again we assume continuity, i.e.\ $Y\left(s_{2L}\right)=Y\left(s_{2R}\right)$. Now we can compute the effective admittance at $x_1$, $Y_{eff}\left(x_1\right)=\dfrac{-{Q_s}_2(s_{1R},t)}{p_2(s_{1R},t)}$ (notice again the sign change in ${Q_s}_2$) as

\begin{equation}
Y_{eff}\left(x_{1R}\right)=iY(s_{1R})\frac{\ J_\frac{4}{3}\left(z_{1R}\right)+\frac{B_2}{A_2}{\ Y}_\frac{4}{3}\left(z_{1R}\right)}{{\ J}_\frac{1}{3}\left(z_{1R}\right)+\frac{B_2}{A_2}{\ Y}_\frac{1}{3}\left(z_{1R}\right)}
\label{eq:Yeff_x1}
\end{equation}

We apply the same procedure for compartment 1. The ratio $\frac{p_1^b}{p_1^f}$ is found to be
\begin{equation}
\frac{p_1^b}{p_1^f}=\frac{1-\frac{Y_{eff}\left(x_{1R}\right)}{Y_1}}{1+\frac{Y_{eff}\left(x_{1R}\right)}{Y_1}}e^{-i2\omega\frac{x_1-x_0}{c_1}}
\label{eq:pb_over_pf}
\end{equation}
and the effective admittance at $x_0$ is
\begin{equation}
Y_{eff}\left(x_0\right)=Y_1 \frac{1-\frac{p_1^b}{p_1^f}}{1+\frac{p_1^b}{p_1^f}}
\label{eq:Yeff_x0}
\end{equation}

We are now able to complete the analytical solution. At the boundary $x_0$, $p_1^f+p_1^b=P(x_0)=1$ (assuming unit pressure excitation), and taking into account \eqref{eq:Yeff_x0}, we find that the pressure at an arbitrary point $x$ in the 1st compartment is given by,
\begin{equation}
p_1\left(x,t\right)=P\left(x_0\right)\left[{cos\left(\omega\frac{x-x_0}{c_1}\right)-i\frac{Y_{eff}\left(x_0\right)}{Y_1}\sin{\left(\omega\frac{x-x_0}{c_1}\right)}}\right]e^{i\omega t}
\end{equation}
Similarly we can compute the pressure and volume flow rate at every point along the vessel. 

Summarising, in order to evaluate the analytic solution for every $\omega$, we start form the last compartment and proceed backwards. More specifically, we first compute $\frac{B_3}{A_3}$ from \eqref{eq:B3_over_A3}, and then $\frac{Y_{eff}(x_{2R})}{Y(s_{2R})}$ from \eqref{eq:Yeffx2r}, $\frac{B_2}{A_2}$ from \eqref{eq:B2_over_A2}, $\frac{Y_{eff}(x_{1R})}{Y(s_{1R})}$ from \eqref{eq:Yeff_x1}, $\frac{p_1^b}{p_1^f}$ from \eqref{eq:pb_over_pf} and $\frac{Y_eff (x_0)}{Y_1}$ from \eqref{eq:Yeff_x0} in that order. To the best of our knowledge, these analytic solutions have not appeared in the literature before.

The fact that a recursive solution can be obtained is directly related to the lower block diagonal structure of the system matrix, as mentioned in section \ref{sec:1Dmodel}. It is important to note that the same recursive approach can be easily extended to handle the whole cardiovascular tree. Appropriate compatibility conditions (pressure continuity and mass conservation) are applied at the tree junctions, and these allow the computation of the local admittances. More details can be found in \cite{Lighthill_1975}. 

\section{Comparison of the analytic solution with the numerical solution of the full model.}
\label{sec:comparison_analytical_numerical}
\begin{figure}[!htb]
\centering
\begin{subfigure}[b]{0.49\textwidth}
\includegraphics[width=0.99\textwidth]{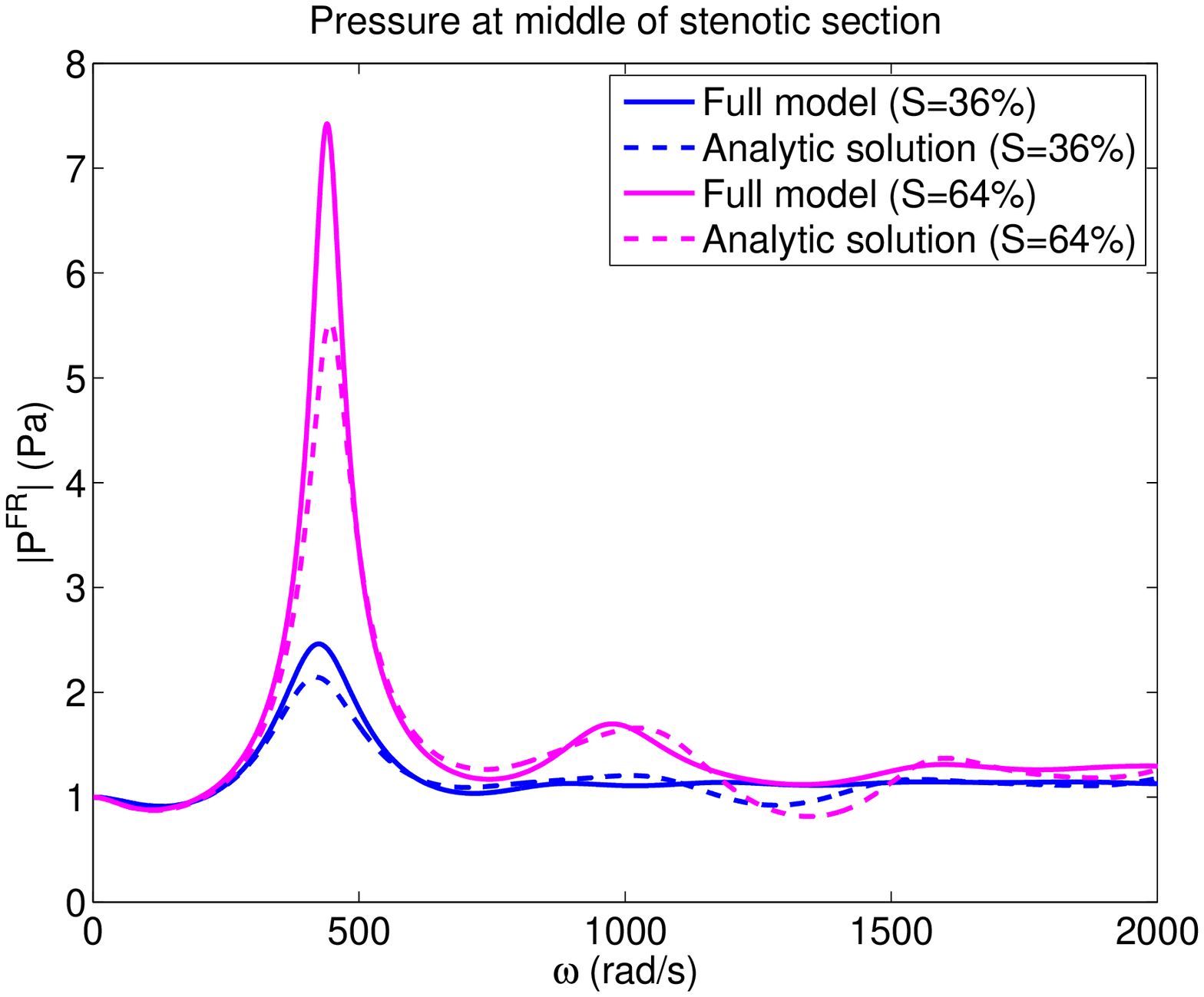}
\caption{}
\end{subfigure}
\begin{subfigure}[b]{0.49\textwidth}
\includegraphics[width=0.99\textwidth]{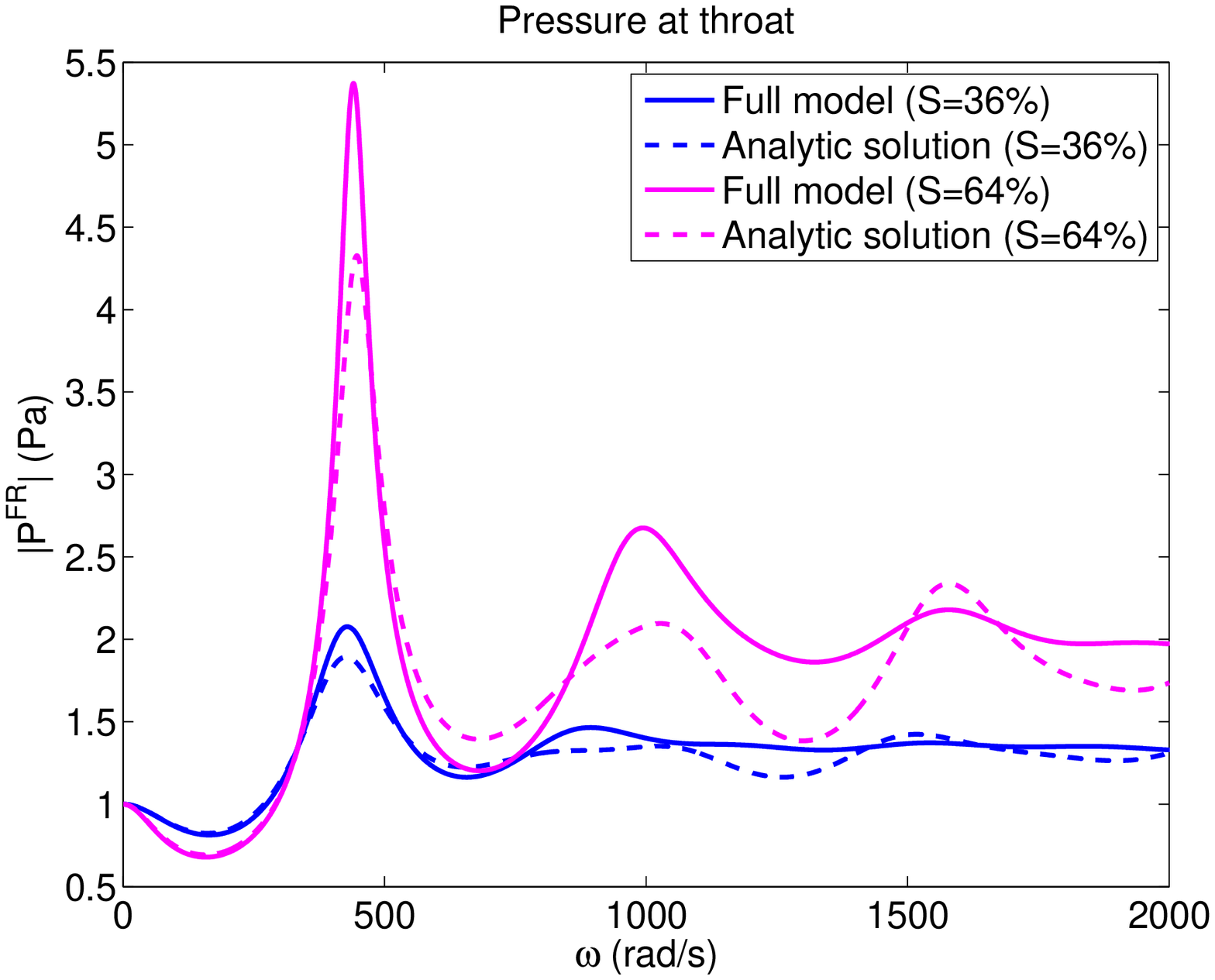}
\caption{}
\end{subfigure}
~
\begin{subfigure}[b]{0.49\textwidth}
\includegraphics[width=0.99\textwidth]{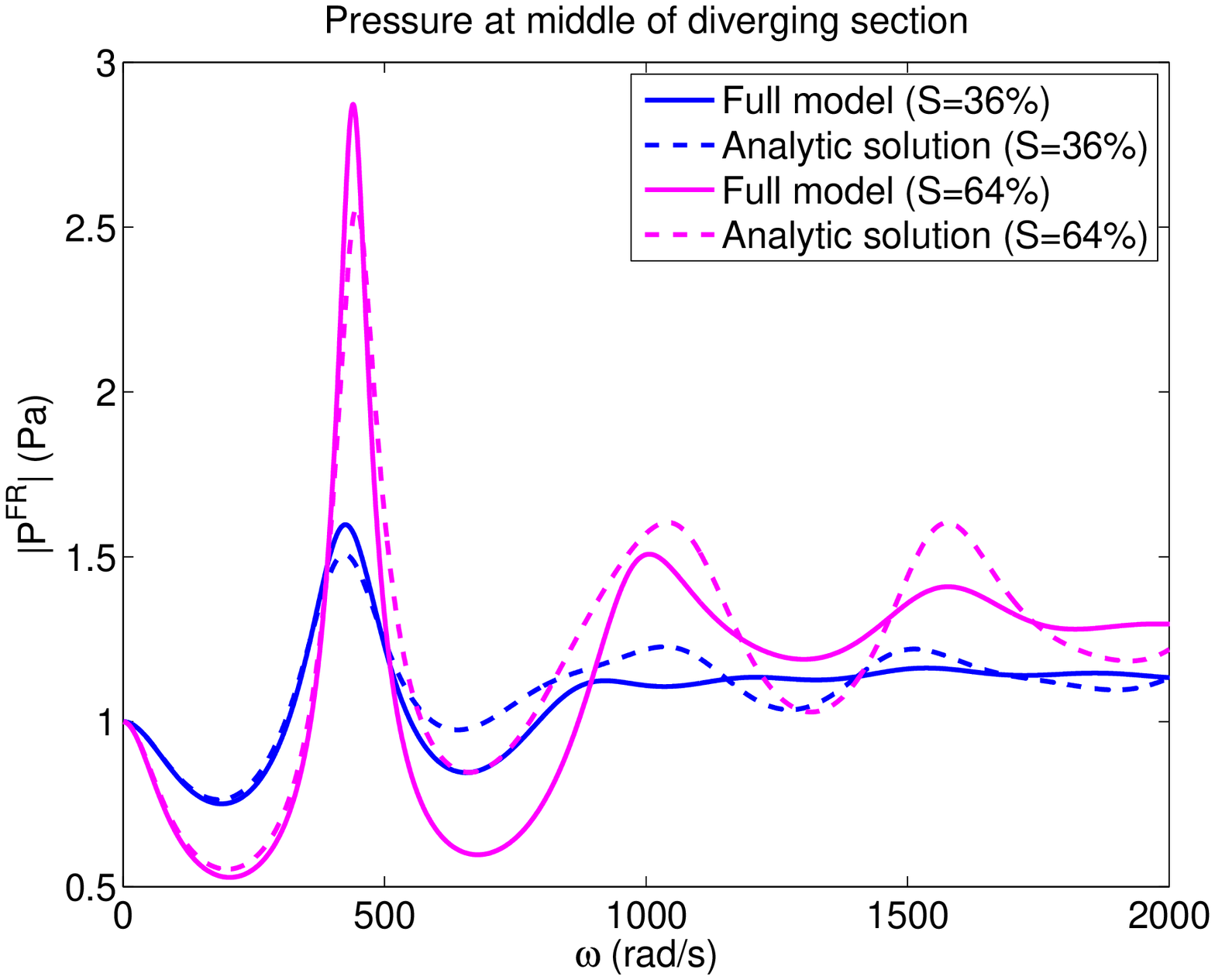}
\caption{}
\end{subfigure}
\begin{subfigure}[b]{0.49\textwidth}
\includegraphics[width=0.99\textwidth]{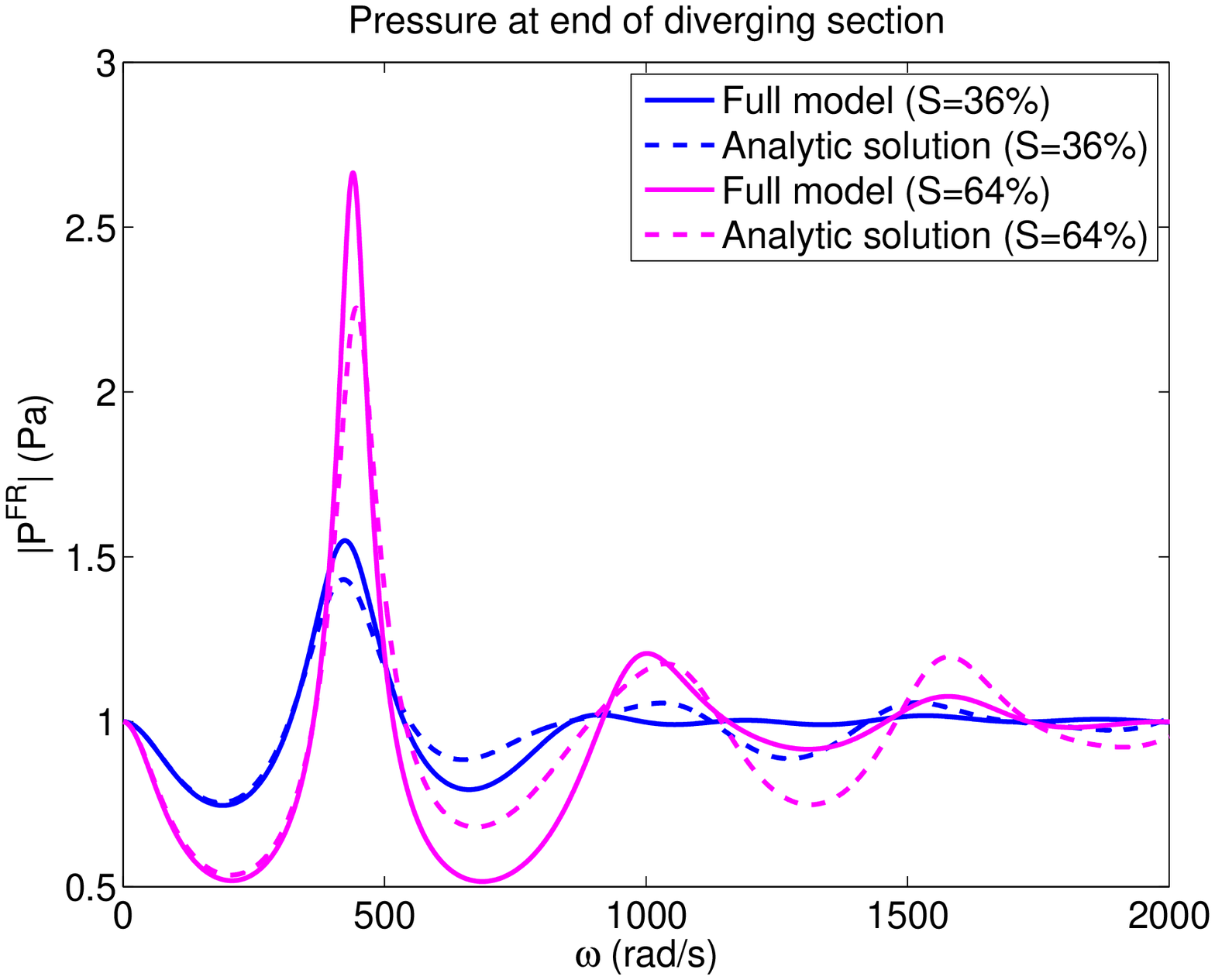}
\caption{}
\end{subfigure}
\caption{Comparison of the pressure frequency response $|P^{FR}(\omega)| (Pa)$ between the full and the analytical solution of the simplified model at 4 locations along the vessel.}
\label{fig:pressure_comparison}
\end{figure}

\begin{figure}[!htb]
\centering
\begin{subfigure}[b]{0.49\textwidth}
\includegraphics[width=0.99\textwidth]{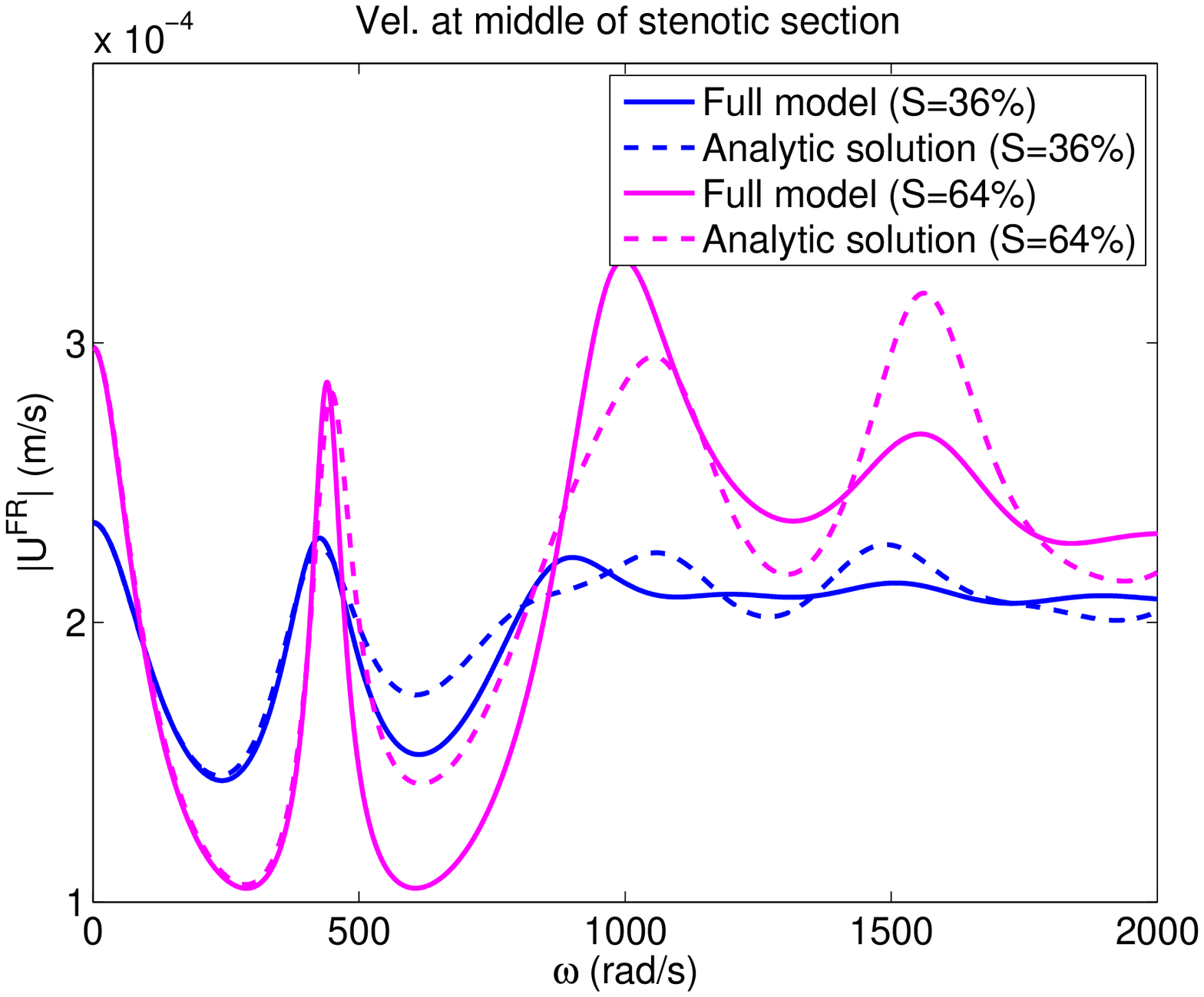}
\caption{}
\end{subfigure}
\begin{subfigure}[b]{0.49\textwidth}
\includegraphics[width=0.99\textwidth]{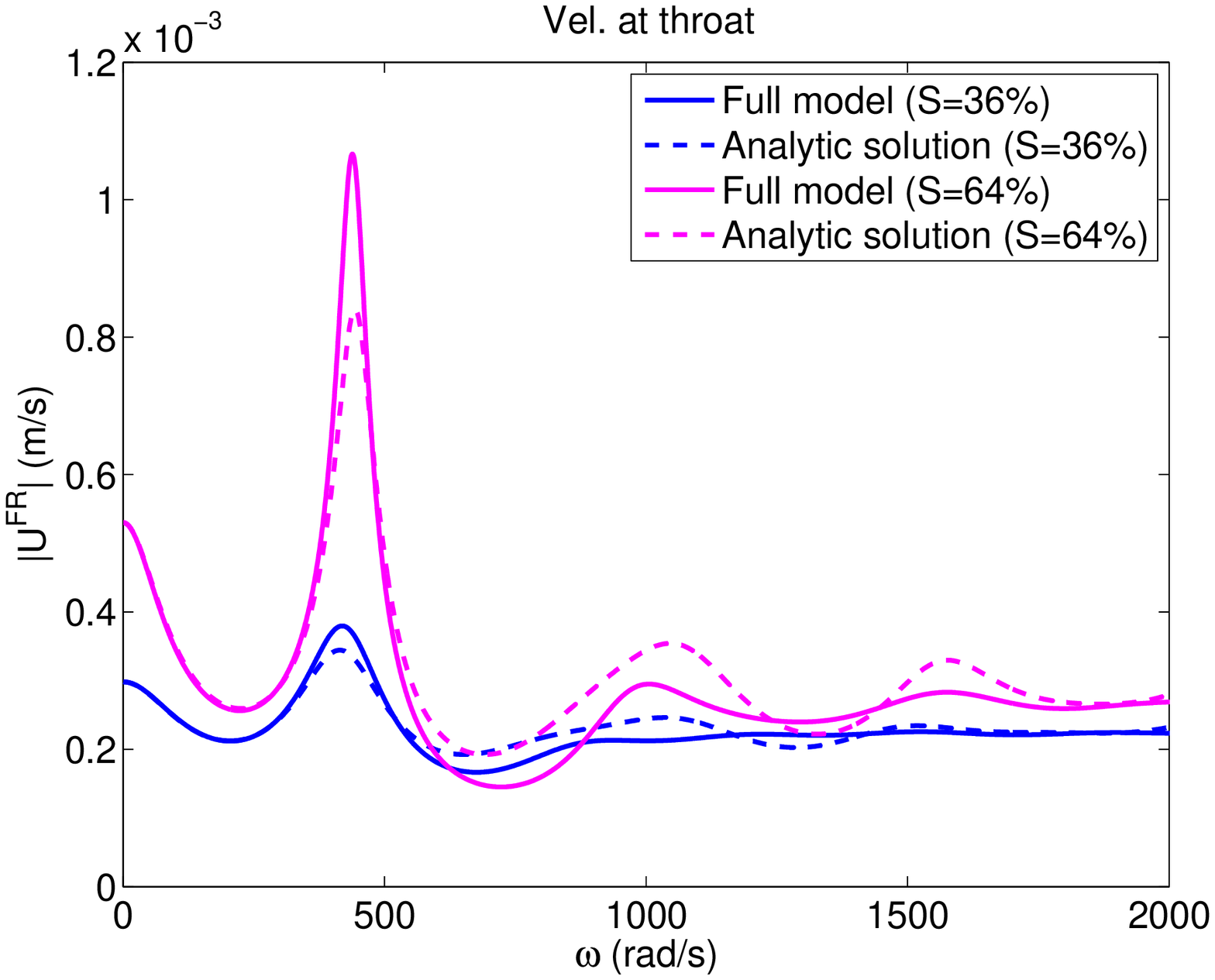}
\caption{}
\end{subfigure}
~
\begin{subfigure}[b]{0.49\textwidth}
\includegraphics[width=0.99\textwidth]{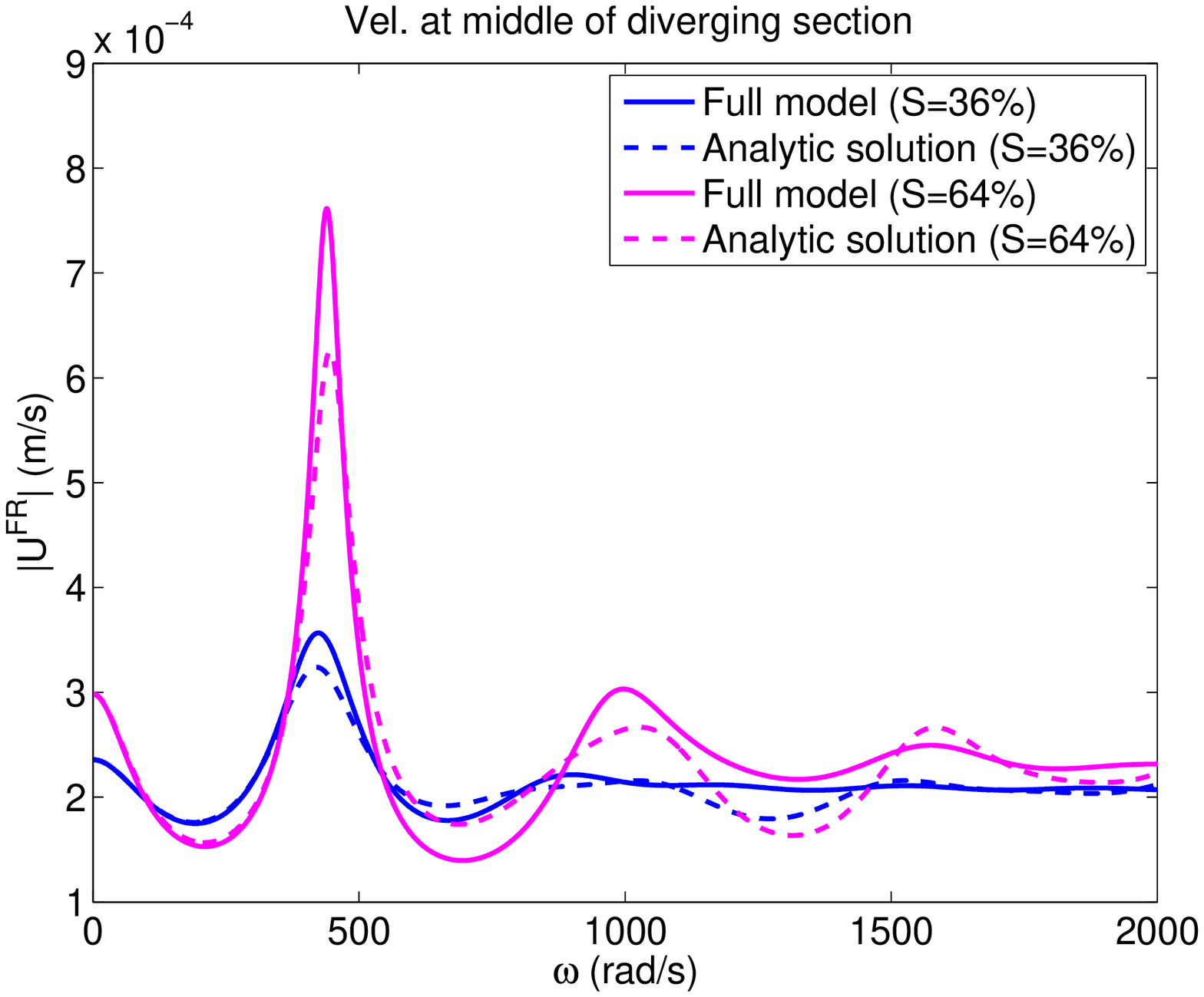}
\caption{}
\end{subfigure}
\begin{subfigure}[b]{0.49\textwidth}
\includegraphics[width=0.99\textwidth]{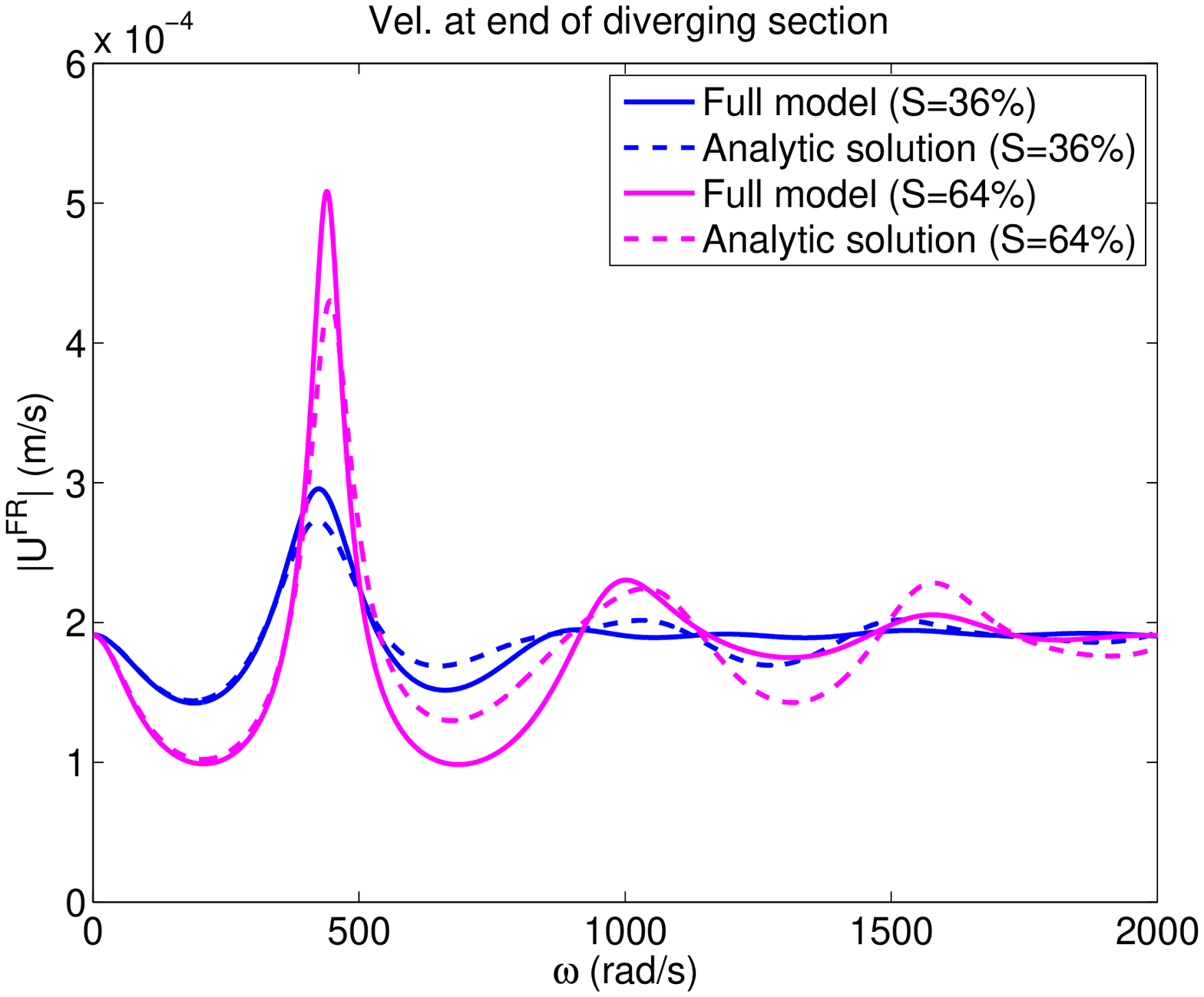}
\caption{}
\end{subfigure}
\caption{Comparison of the velocity frequency response $|U^{FR}(\omega)| (m/s)$ between the full and the analytical solution of the simplified model at 4 locations along the vessel.}
\label{fig:velocity_comparison}
\end{figure}

The frequency response of the full model is compared with the analytic solution obtained in the previous section in figures \ref{fig:pressure_comparison} and \ref{fig:velocity_comparison} for pressure and velocity respectively.

It is very interesting to see that the simplified model can capture very accurately the frequency response of the full model for both pressure and velocity at all locations in a frequency range less than about $~400 rad/s$. This range is well within the physiological regime. The resonance frequency is also very well predicted, but the value is slightly underestimated. 

These results indicate that only two parameters, namely the length and degree of stenosis, suffice to capture the behaviour of the vessel at the relevant frequencies. The exact details of the geometry affect the magnitude of the resonance peak and the response at large frequencies only.  

We can easily explain this behaviour. When the frequency of the input signal is small, the wavelength is large and so the response is determined by the large-scale features of the stenotic vessel (degree of stenosis and length). On the other hand, larger frequencies that correspond to shorter wavelengths, are affected by the finer details of vessel, for example the exact stenotic profile. Of course, this is within the context of linear, inviscid theory. The effect of viscosity will be examined in the next section.  

\section{Comparison between 1D and 3D models}\label{sec:comparison_3D_1D}
To assess the performance of the 1D model, we compare against the results of a 3D FSI model of an idealised stenotic carotid artery (see \cite{Figueroa_et_al_2006}). The 3D model employs a membrane formulation for the vessel wall, which is considered as a linear elastic material. The membrane formulation, without transverse shear, was also employed to derive equations \eqref{eq:continuity_momentum_sph} of the 1D model. The fluid and structure equations are solved using the coupled momentum (or transpiration) method on a fixed computational domain. Discretisation is based on the finite element methods for both media.

The physical parameters are provided in table \ref{tab:FSI_vessel_parameters}. The throat is located at $2.5cm$ from the inlet and the total length of the vessel is $12.6cm$. The degree of stenosis is 75\% and the ratio of stenotic length to inlet diameter, $L/D=1.67$.

\begin{table}[!htb]
\centering
\begin{tabular}{|l|l|}
\hline
Modulus of Elasticity ($dyn/cm^2$) &	$4.07\times 10^6$ \\
\hline
Fluid density ($g/cm^3$) &	$1.06$ \\
\hline
Membrane density ($g/cm^3$) &	$1.0$ \\
\hline
Membrane thickness ($cm$) &	$0.03$ \\
\hline
Poisson ratio	 & $0.5$ \\
\hline
Inlet/Outlet vessel diameter ($cm$) & $0.6$ \\
\hline
Throat diameter ($cm$) & $0.3$ \\
\hline
Length of stenotic region ($cm$) & $1.0$ \\
\hline
\end{tabular}
\caption{Fluid and membrane parameters of the 3D FSI model (data taken from \cite{Figueroa_et_al_2006}).}
\label{tab:FSI_vessel_parameters}
\end{table}

At the inlet, the volume flow rate is imposed (shown as the blue line in figure \ref{fig:Taylor_comparison_volume}). During the cardiac cycle, the Reynolds number based on the inlet diameter, $D$, and the cross-section average velocity, $u_m(t)$, varies between $213$ and $767$, the time-average being $365$ (the peak-to-mean ratio is $3.6$). The reduced velocity $U_{red}= \overline{u_m(t)}T/D=42$, where the overbar denotes time- averaging and $T=1.1s$ is the period of the cardiac cycle. Under these conditions, the flow downstream of the throat was found to be transitional and non-axisymmetric. This is consistent with the results of stability analysis of a stenotic vessel with the same $S$ but slightly larger $L/D$ ($2$ instead of $1.67$) of \cite{sherwin_blackburn_2005}. For steady inlet flow, the authors observed transitional flow for Reynolds larger than $688$ (for our case, the $Re$ exceeds this value for part of the cardiac cycle). The transition was found to be due a sub-critical bifurcation that leads to loss of axisymmetry and a weak Coanda-type deflection of the jet emanating from the throat towards the walls (similar behavior is shown in Figure 7 of \cite{Figueroa_et_al_2006}). \cite{sherwin_blackburn_2005} also considered transition due to pulsating inlet flow, but their reduced velocities are much smaller (up to 7.5) compared to the present case. It is therefore more appropriate to compare with the steady results, as the reduced velocity is quite large. 

In order to apply the 1D model, the flow rate is decomposed in Fourier series, with fundamental period $T$. Most of the energy is included in the fundamental mode and the first 10 harmonics. The response for each harmonic was computed in the frequency domain using the approach of section \ref{sec:1Dmodel} and the result was converted back to the time domain using inverse Fourier transform. The steady component was treated separately; the time-average volume flow rate was imposed at the inlet and the velocity at each location was computed using the continuity equation. 

An impedance boundary condition was imposed at the outlet. More specifically, the volume and pressure signals at the outlet were written in Fourier series and the complex impedance $Z(\omega_i)=P(\omega_i)/Q(\omega_i)$ was computed for each frequency $\omega_i$. The imposed boundary condition was then $P(\omega_i)=Z(\omega_i) A U_s(\omega_i)$, where $A=\pi D^2/4$ is the outlet area.  This approach reproduces the exact boundary condition of the 3D FSI model and eliminates any uncertainties in the comparison that can arise from inconsistent outlet conditions.

\begin{figure}[!htb]
\centering
\begin{subfigure}[b]{0.70\textwidth}
\includegraphics[width=0.99\textwidth]{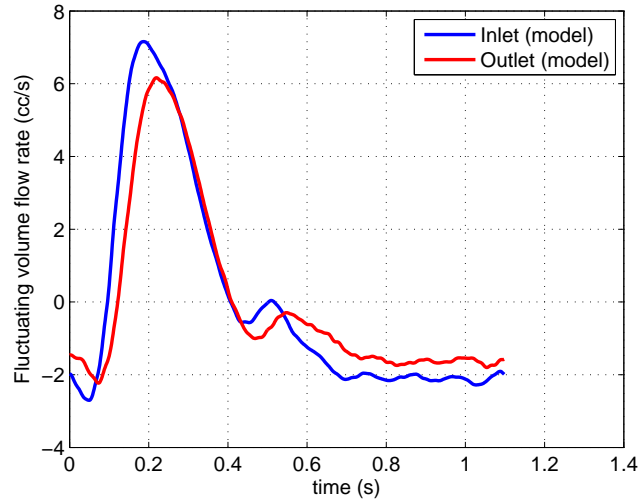}
\caption{}
\end{subfigure}
\begin{subfigure}[b]{0.70\textwidth}
\includegraphics[width=0.99\textwidth]{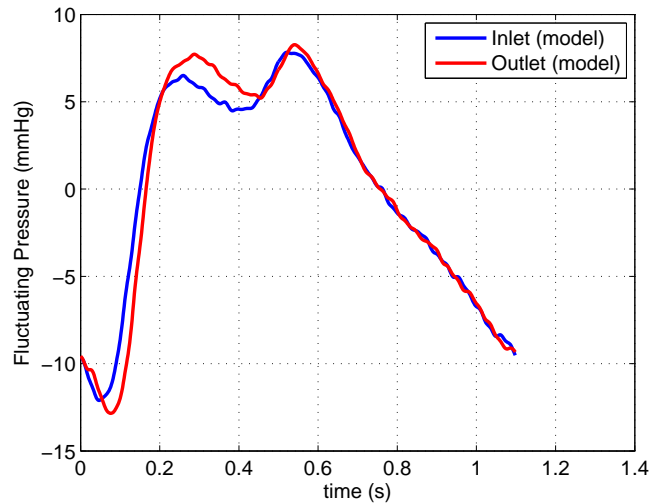}
\caption{}
\label{fig:Taylor_fluctuating_pressure}
\end{subfigure}
\caption{Fluctuating (a) volume flow rate and (b) pressure at the inlet and outlet of the vessel as predicted by the 1D model.}
\label{fig:Taylor_fluctuating}
\end{figure}

The fluctuating volume flow rate and pressure at the inlet and outlet of the domain during the entire cardiac cycle are shown in figure \ref{fig:Taylor_fluctuating}. There is a phase difference between the inlet and outlet due to the finite propagation velocity. This is not constant during systole because of the variation of the propagation velocity due to the presence of stenosis. Note also that during diastole, the volume flow rate at the outlet is larger compared to the inlet. This is because volume has accumulated inside the flexible vessel during systole, and released during diastole. This behaviour can only be captured using a model that accounts for vessel distensibility.  

\begin{figure}[!htb]
\centering
\begin{subfigure}[b]{0.70\textwidth}
\includegraphics[width=0.99\textwidth]{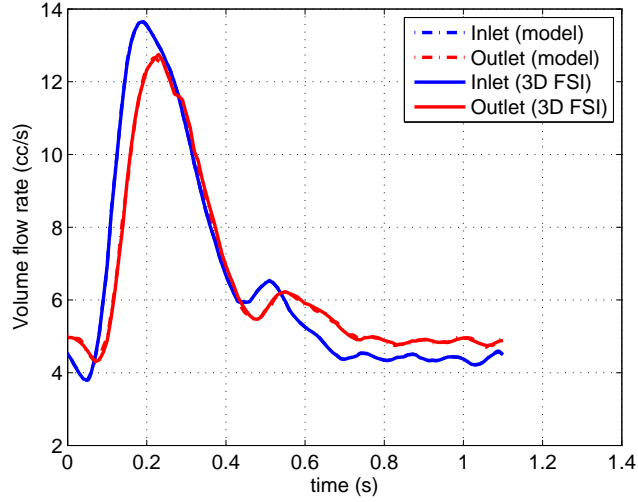}
\caption{}
\label{fig:Taylor_comparison_volume}
\end{subfigure}
\begin{subfigure}[b]{0.70\textwidth}
\includegraphics[width=0.99\textwidth]{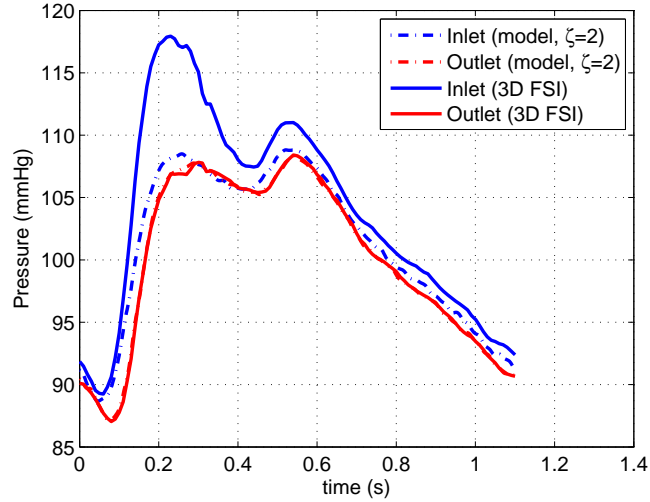}
\caption{}
\label{fig:Taylor_comparison_pressure_zeta_2}
\end{subfigure}
\caption{Comparison between 1D and 3D FSI models for (a) volume flow rate and (b) pressure at the inlet and outlet of the vessel (for constant $\zeta=2$ in the whole vessel).}
\label{fig:Taylor_comparison}
\end{figure}

Adding the steady component of velocity to the fluctuating part, we obtain the total volume flow rate, which is compared to the 3D results in figure \ref{fig:Taylor_comparison_volume}. At the outlet the values predicted by the 1D model match very closely with the 3D values during the entire periodic cycle.  Indeed, the error between the 1D and 3D models (dashed and solid red lines respectively) is less than $1\%$. Note that phase difference between inlet and outlet mentioned above is captured very accurately.

The computation of pressure is more involved. In the momentum equation of the 1D model we have ignored the non-linear and the viscous terms. Both are expected to be significant in the present case. In order to take them into account approximately, we write the 1D momentum equation as,
\begin{equation}
\frac{\partial U_x}{\partial t}+ U_x \frac{\partial U_x}{\partial x} =  -\frac{1}{\rho}\frac{\partial P}{\partial x} + \frac{f}{\rho A}
\label{eq:momentum_more_accurate}
\end{equation}

\noindent where $f(x,t)$ is the frictional force per unit length and we have assumed that the profile shape factor is equal to $1$ (for details refer to \cite{Alastruey_2012}). Assuming the velocity profile,
\begin{equation}
u(x,r,t)=U_x(x,t)\frac{\zeta+2}{\zeta} \left[ 1- \left ( \frac {r}{R} \right)^{\zeta} \right]
\label{eq:velocity_profile}
\end{equation}

\noindent that satisfies the non-slip condition $u(x,R,t)=0$ at the wall,  we find that $f(x,t)=-2(\zeta+2)\mu \pi U_x(x,t)$, where $\mu$ is the dynamic viscosity. Parameter $\zeta$ determines the shape of the profile. The standard parabolic profile corresponds to $\zeta=2$. Increasing $\zeta$ makes the profile flatter around the centerline and increases shear close to the wall. Using the values of $U_x(x,t)$ from the 1D inviscid model (that were shown to match well the 3D results) and substituting in \eqref{eq:momentum_more_accurate}, we obtain the pressure gradient $\frac{\partial P}{\partial x}$. In order to compute $P(x,t)$, a boundary condition is needed and a value of $\zeta$.  Using the time-average value of $P$ from the 3D results at the outlet and adding the fluctuating pressure shown in figure \ref{fig:Taylor_fluctuating_pressure} we obtain the required boundary condition. At the outlet the volume flow rate is well predicted,  
and since a boundary condition that matches the impedance between the 3D and 1D model was applied, we expect that the pressure profile will be also well predicted. 

\begin{figure}[!htb]
\centering
\includegraphics[width=0.70\textwidth]{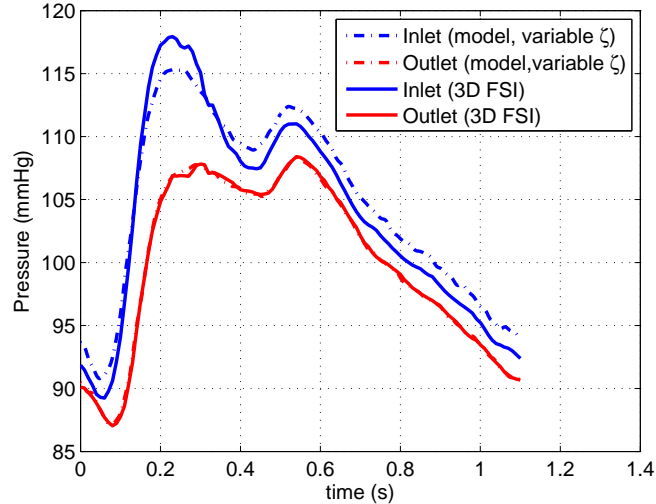}
\caption{Comparison between 1D and 3D FSI models for pressure at the inlet and outlet of the vessel (for $\zeta=44$ at stenosis region and $2$ elsewhere).}
\label{fig:Taylor_comparison_variable_zeta}
\end{figure}

This is indeed the case, as can be clearly seen in figure \ref{fig:Taylor_comparison_pressure_zeta_2} (red solid and dashed lines overlap). This confirms that our implementation is correct. The results in this figure were obtained with $\zeta=2$ for the whole domain. The inlet pressure however is severely under-predicted because the pressure losses in the stenotic region are not accounted for correctly. The value of $\zeta=2$ corresponds to a Poisseuille profile while the DNS results of \cite{varghese_frankel_fischer_2007a, varghese_frankel_fischer_2007b} indicate that in the stenosis the profile becomes flatter, leading to much higher shear stress. For example, for steady inlet $Re=500$, the wall shear stress varies very sharply across the stenotic region (refer to figure 18 of \cite{varghese_frankel_fischer_2007a}) and at the throat is $22$ times larger compared to upstream levels. Interestingly, the location of maximum shear amplification factor (equal to 30) is immediately prior of the throat. At $Re=1000$, the maximum values increase by more than a factor of 40. 

Based on this evidence from DNS results, we performed an additional computation in which $\zeta=44$ in the stenotic region and $2$ everywhere else. The large value of $\zeta$ corresponds to an average increase of the shear by $11.5$ times compared to Poisseuille profile. The results are shown in figure \ref{fig:Taylor_comparison_variable_zeta}. It is clear that the peak value of pressure is now much better approximated but the secondary peak and the values during systole are slightly overpredicted.  As $\zeta$ depends strongly on $Re$, it is expected that more accurate predictions could have been obtained if a time- and space-dependent $\zeta$ is considered. However we have not pursued this further as the current results can be  considered acceptable; for example the error at the peak is less than $3\%$.
  
\section{Conclusions}\label{sec:conclusion}
An 1D model to represent stenotic vessels has been proposed. Unlike models previously developed in the literature, the vessel is composed of series of tapered sections, instead of sections with constant area. At each section, the pressure and velocity can be expressed analytically in terms of Bessel functions. Using appropriate compatibility and boundary conditions between the sections, the full 1D pressure and velocity fields can be quickly obtained numerically in the frequency domain. The superposition of forward and reflecting waves results at resonant frequencies. 

By representing the stenosis using only two segments, we can derive closed form analytical solutions that match very well the numerical solution at small, physiologically relevant, frequencies. The 1D model results were compared against 3D solutions that account for fluid-structure-interaction. Excellent matching was achieved between for the volume flow rate, but differences were noted on the pressure distribution. Using input from DNS simulations to approximate better the velocity profile around the stenotic region, significantly better results were obtained. 

The proposed method can be used to approximate arbitrary profile shapes. It can also be extended to handle the whole cardiovascular tree. This will make sensitivity and uncertainty quantification studies much more affordable than they currently are.   

\clearpage

%References
%\textbf{References}
\bibliography{References_MD}

\begin{thebibliography}{29}
\expandafter\ifx\csname natexlab\endcsname\relax\def\natexlab#1{#1}\fi
\providecommand{\url}[1]{\texttt{#1}}
\providecommand{\href}[2]{#2}
\providecommand{\path}[1]{#1}
\providecommand{\DOIprefix}{doi:}
\providecommand{\ArXivprefix}{arXiv:}
\providecommand{\URLprefix}{URL: }
\providecommand{\Pubmedprefix}{pmid:}
\providecommand{\doi}[1]{\href{http://dx.doi.org/#1}{\path{#1}}}
\providecommand{\Pubmed}[1]{\href{pmid:#1}{\path{#1}}}
\providecommand{\bibinfo}[2]{#2}
\ifx\xfnm\relax \def\xfnm[#1]{\unskip,\space#1}\fi
%Type = Article
\bibitem[{Ahmed and Giddens(1983)}]{Ahmed_Giddens_1983a}
\bibinfo{author}{Ahmed, S.A.}, \bibinfo{author}{Giddens, D.P.},
  \bibinfo{year}{1983}.
\newblock \bibinfo{title}{Velocity measurements in steady flow through
  axisymmetric stenoses at moderate {Reynolds} number}.
\newblock \bibinfo{journal}{Journal of Biomechanics} \bibinfo{volume}{16(7)},
  \bibinfo{pages}{505–--516}.
%Type = Article
\bibitem[{Alastruey et~al.(2011)Alastruey, Khir, Matthys, Segers, Sherwin,
  Verdonck, Parker and Peiró}]{Alastruey_et_al_2011}
\bibinfo{author}{Alastruey, J.}, \bibinfo{author}{Khir, A.W.},
  \bibinfo{author}{Matthys, K.S.}, \bibinfo{author}{Segers, P.},
  \bibinfo{author}{Sherwin, S.}, \bibinfo{author}{Verdonck, P.},
  \bibinfo{author}{Parker, K.}, \bibinfo{author}{Peiró, J.},
  \bibinfo{year}{2011}.
\newblock \bibinfo{title}{Pulse wave propagation in a model human arterial
  network: Assessment of {1-D} visco-elastic simulations against in vitro
  measurements}.
\newblock \bibinfo{journal}{Journal of Biomechanics} \bibinfo{volume}{44(12)},
  \bibinfo{pages}{2250--2258}.
%Type = Incollection
\bibitem[{Alastruey et~al.(2012)Alastruey, Parker and Sherwin}]{Alastruey_2012}
\bibinfo{author}{Alastruey, J.}, \bibinfo{author}{Parker, K.H.},
  \bibinfo{author}{Sherwin, S.J.}, \bibinfo{year}{2012}.
\newblock \bibinfo{title}{Arterial pulse wave haemodynamics}, in:
  \bibinfo{booktitle}{11th International Conference on Pressure Surges}.
  \bibinfo{publisher}{Virtual PiE Led t/a BHR Group}, pp.
  \bibinfo{pages}{401--443}.
%Type = Article
\bibitem[{Berger and Jou(2000)}]{Berger_Jou_2000}
\bibinfo{author}{Berger, S.}, \bibinfo{author}{Jou, L.D.},
  \bibinfo{year}{2000}.
\newblock \bibinfo{title}{Flows in stenotic vessels}.
\newblock \bibinfo{journal}{Annual Review of Fluid Mechanics}
  \bibinfo{volume}{32}, \bibinfo{pages}{347--382}.
%Type = Article
\bibitem[{Cassanova and Giddens(1978)}]{Cassanova_Giddens_1978}
\bibinfo{author}{Cassanova, R.A.}, \bibinfo{author}{Giddens, D.P.},
  \bibinfo{year}{1978}.
\newblock \bibinfo{title}{Disorder distal to modeled stenoses in steady and
  pulsatile flow}.
\newblock \bibinfo{journal}{Journal of Biomechanics} \bibinfo{volume}{11},
  \bibinfo{pages}{441--453}.
%Type = Article
\bibitem[{Chen et~al.(2013)Chen, Quarteroni and Rozza}]{Chen_et_al_2013}
\bibinfo{author}{Chen, P.}, \bibinfo{author}{Quarteroni, A.},
  \bibinfo{author}{Rozza, G.}, \bibinfo{year}{2013}.
\newblock \bibinfo{title}{Simulation-based uncertainty quantification of human
  arterial network hemodynamics}.
\newblock \bibinfo{journal}{International Journal for Numerical Methods in
  Biomedical Engineering} \bibinfo{volume}{29}, \bibinfo{pages}{698--721}.
%Type = Article
\bibitem[{Choi et~al.(2017)Choi, Park, Huh and S.J.}]{Choi_et_al_2017}
\bibinfo{author}{Choi, W.}, \bibinfo{author}{Park, S.}, \bibinfo{author}{Huh,
  H.}, \bibinfo{author}{S.J., L.}, \bibinfo{year}{2017}.
\newblock \bibinfo{title}{Hemodynamic characteristics of flow around a
  deformable stenosis}.
\newblock \bibinfo{journal}{Journal of Biomechanics} \bibinfo{volume}{61},
  \bibinfo{pages}{216--223}.
%Type = Article
\bibitem[{Du et~al.(2016)Du, Hu and Cai}]{Du_et_al_2016}
\bibinfo{author}{Du, T.}, \bibinfo{author}{Hu, D.}, \bibinfo{author}{Cai, D.},
  \bibinfo{year}{2016}.
\newblock \bibinfo{title}{A fast algorithm for the simulation of arterial pulse
  waves}.
\newblock \bibinfo{journal}{Journal of Computational Physics}
  \bibinfo{volume}{314}, \bibinfo{pages}{450--464}.
%Type = Article
\bibitem[{Figueroa et~al.(2006)Figueroa, Vignon-Clementel, Jansen, Hughes and
  Taylor}]{Figueroa_et_al_2006}
\bibinfo{author}{Figueroa, C.}, \bibinfo{author}{Vignon-Clementel, I.},
  \bibinfo{author}{Jansen, K.}, \bibinfo{author}{Hughes, T.},
  \bibinfo{author}{Taylor, C.}, \bibinfo{year}{2006}.
\newblock \bibinfo{title}{A coupled momentum method for modeling blood flow in
  three-dimensional deformable arteries}.
\newblock \bibinfo{journal}{Computer Methods in Applied Mechanics and
  Engineering} \bibinfo{volume}{195 (41)}, \bibinfo{pages}{5685--5706}.
%Type = Article
\bibitem[{Flores et~al.(2016)Flores, Alastruey and
  Corvera~Poiré}]{Flores_et_al_2016}
\bibinfo{author}{Flores, J.}, \bibinfo{author}{Alastruey, J.},
  \bibinfo{author}{Corvera~Poiré, E.}, \bibinfo{year}{2016}.
\newblock \bibinfo{title}{A novel analytical approach to pulsatile blood flow
  in the arterial network}.
\newblock \bibinfo{journal}{Annals of Biomedical Engineering}
  \bibinfo{volume}{44(10)}, \bibinfo{pages}{3047--3068}.
%Type = Article
\bibitem[{Gerbeau et~al.(2005)Gerbeau, Vidrascu and Frey}]{Gerbeau_et_al_2005}
\bibinfo{author}{Gerbeau, J.F.}, \bibinfo{author}{Vidrascu, M.},
  \bibinfo{author}{Frey, P.}, \bibinfo{year}{2005}.
\newblock \bibinfo{title}{Fluid–structure interaction in blood flows on
  geometries based on medical imaging}.
\newblock \bibinfo{journal}{Computers and Structures} \bibinfo{volume}{83},
  \bibinfo{pages}{155--165}.
%Type = Article
\bibitem[{Huberts et~al.(2013)Huberts, de~Jonge, van~der Linden, Inda, Tordoir,
  van~de Vosse and Bosboom}]{Huberts_al_2013}
\bibinfo{author}{Huberts, W.}, \bibinfo{author}{de~Jonge, C.},
  \bibinfo{author}{van~der Linden, W.}, \bibinfo{author}{Inda, M.},
  \bibinfo{author}{Tordoir, J.}, \bibinfo{author}{van~de Vosse, F.},
  \bibinfo{author}{Bosboom, E.}, \bibinfo{year}{2013}.
\newblock \bibinfo{title}{A sensitivity analysis of a personalized pulse wave
  propagation model for arteriovenous fistula surgery. {Part A}: identification
  of most influential model parameters.}
\newblock \bibinfo{journal}{Medical Engineering and Physics}
  \bibinfo{volume}{35}, \bibinfo{pages}{810--826}.
%Type = Article
\bibitem[{Karri and Vlachos(2010)}]{Karri_Vlachos_2010}
\bibinfo{author}{Karri, S.}, \bibinfo{author}{Vlachos, P.},
  \bibinfo{year}{2010}.
\newblock \bibinfo{title}{Time-resolved {DPIV} investigation of pulsatile flow
  in symmetric stenotic arteries—effects of phase angle}.
\newblock \bibinfo{journal}{ASME Journal of Biomechanical Engineering}
  \bibinfo{volume}{132(3)}, \bibinfo{pages}{031010}.
%Type = Article
\bibitem[{Ku(1997)}]{Ku_1997}
\bibinfo{author}{Ku, D.}, \bibinfo{year}{1997}.
\newblock \bibinfo{title}{Blood flow in arteries}.
\newblock \bibinfo{journal}{Annual Review of Fluid Mechanics}
  \bibinfo{volume}{29}, \bibinfo{pages}{399--434}.
%Type = Incollection
\bibitem[{Lighthill(1975)}]{Lighthill_1975}
\bibinfo{author}{Lighthill, J.}, \bibinfo{year}{1975}.
\newblock \bibinfo{title}{Pulse propagation theory}, in:
  \bibinfo{booktitle}{Mathematical Biofluid Dynamics}.
  \bibinfo{publisher}{CBMS-NSF Regional Conference Series in Applied
  Mathematics, vol. 17. SIAM}. chapter~\bibinfo{chapter}{12}, pp.
  \bibinfo{pages}{227--252}.
%Type = Article
\bibitem[{Mynard and Nithiarasu(2008)}]{Mynard_Nithiarasu_2008}
\bibinfo{author}{Mynard, J.P.}, \bibinfo{author}{Nithiarasu, P.},
  \bibinfo{year}{2008}.
\newblock \bibinfo{title}{A {1D} arterial blood flow model incorporating
  ventricular pressure, aortic valve and regional coronary flow using the
  locally conservative galerkin ({LCG}) method}.
\newblock \bibinfo{journal}{Commun. Numer. Meth. Engng} \bibinfo{volume}{24},
  \bibinfo{pages}{367–--417}.
%Type = Article
\bibitem[{Ojha et~al.(1989)Ojha, Cobbold, Johnston and
  Hummel}]{ojha_cobbold_johnston_hummel_1989}
\bibinfo{author}{Ojha, M.}, \bibinfo{author}{Cobbold, R.S.C.},
  \bibinfo{author}{Johnston, K.W.}, \bibinfo{author}{Hummel, R.L.},
  \bibinfo{year}{1989}.
\newblock \bibinfo{title}{Pulsatile flow through constricted tubes: an
  experimental investigation using photochromic tracer methods}.
\newblock \bibinfo{journal}{Journal of Fluid Mechanics} \bibinfo{volume}{203},
  \bibinfo{pages}{173--197}.
%Type = Article
\bibitem[{Pal et~al.(2014)Pal, Anupindi, Delorme, Ghaisas, Shetty and
  Frankel}]{Pal_et_al_2014}
\bibinfo{author}{Pal, A.}, \bibinfo{author}{Anupindi, K.},
  \bibinfo{author}{Delorme, Y.}, \bibinfo{author}{Ghaisas, N.},
  \bibinfo{author}{Shetty, D.}, \bibinfo{author}{Frankel, S.},
  \bibinfo{year}{2014}.
\newblock \bibinfo{title}{{Large} {Eddy} {Simulation} of transitional flow in
  an idealized stenotic blood vessel: Evaluation of subgrid scale models}.
\newblock \bibinfo{journal}{ASME Journal of Biomechanical Engineering}
  \bibinfo{volume}{136(7)}, \bibinfo{pages}{071009}.
%Type = Article
\bibitem[{Papadakis(2011)}]{Papadakis_2011}
\bibinfo{author}{Papadakis, G.}, \bibinfo{year}{2011}.
\newblock \bibinfo{title}{New analytic solutions for wave propagation in
  flexible, tapered vessels with reference to mammalian arteries}.
\newblock \bibinfo{journal}{Journal of Fluid Mechanics} \bibinfo{volume}{689},
  \bibinfo{pages}{465--488}.
%Type = Article
\bibitem[{Pielhop et~al.(2012)Pielhop, Klaas and
  Schr{\"o}der}]{PIELHOP_et_al_2012}
\bibinfo{author}{Pielhop, K.}, \bibinfo{author}{Klaas, M.},
  \bibinfo{author}{Schr{\"o}der, M.}, \bibinfo{year}{2012}.
\newblock \bibinfo{title}{Analysis of the unsteady flow in an elastic stenotic
  vessel}.
\newblock \bibinfo{journal}{European Journal of Mechanics - B/Fluids}
  \bibinfo{volume}{35}, \bibinfo{pages}{102--110}.
%Type = Article
\bibitem[{Sherwin et~al.(2003)Sherwin, Franke, Peiró and
  Parker}]{Sherwin_et_al_2003}
\bibinfo{author}{Sherwin, S.}, \bibinfo{author}{Franke, V.},
  \bibinfo{author}{Peiró, J.}, \bibinfo{author}{Parker, K.},
  \bibinfo{year}{2003}.
\newblock \bibinfo{title}{One-dimensional modelling of a vascular network in
  space-time variables}.
\newblock \bibinfo{journal}{Journal of Engineering Mathematics}
  \bibinfo{volume}{47(3-4)}, \bibinfo{pages}{217--250}.
%Type = Article
\bibitem[{Sherwin and Blackburn(2005)}]{sherwin_blackburn_2005}
\bibinfo{author}{Sherwin, S.J.}, \bibinfo{author}{Blackburn, H.M.},
  \bibinfo{year}{2005}.
\newblock \bibinfo{title}{Three-dimensional instabilities and transition of
  steady and pulsatile axisymmetric stenotic flows}.
\newblock \bibinfo{journal}{Journal of Fluid Mechanics} \bibinfo{volume}{533},
  \bibinfo{pages}{297--–327}.
%Type = Article
\bibitem[{Taylor and Draney(2004)}]{Taylor_Draney_2004}
\bibinfo{author}{Taylor, C.A.}, \bibinfo{author}{Draney, M.T.},
  \bibinfo{year}{2004}.
\newblock \bibinfo{title}{Experimental and computational methods in
  cardiovascular fluid mechanics}.
\newblock \bibinfo{journal}{Annual Review of Fluid Mechanics}
  \bibinfo{volume}{36}, \bibinfo{pages}{197--231}.
%Type = Article
\bibitem[{{\"U}nsal et~al.(2006){\"U}nsal, Erden and Erden}]{Unsal_et_al_2006}
\bibinfo{author}{{\"U}nsal, N.H.}, \bibinfo{author}{Erden, A.},
  \bibinfo{author}{Erden, I.}, \bibinfo{year}{2006}.
\newblock \bibinfo{title}{Evaluation of the splenic vein diameter and
  longitudinal size of the spleen in patients with {Gamna}-{Gandy} bodies}.
\newblock \bibinfo{journal}{Diagnostic and Interventional Radiology}
  \bibinfo{volume}{12(3)}, \bibinfo{pages}{125--128}.
%Type = Article
\bibitem[{Varghese et~al.(2007a)Varghese, Frankel and
  Fischer}]{varghese_frankel_fischer_2007a}
\bibinfo{author}{Varghese, S.S.}, \bibinfo{author}{Frankel, S.H.},
  \bibinfo{author}{Fischer, P.F.}, \bibinfo{year}{2007}a.
\newblock \bibinfo{title}{Direct numerical simulation of stenotic flows. {Part}
  1. {Steady} flow}.
\newblock \bibinfo{journal}{Journal of Fluid Mechanics} \bibinfo{volume}{582},
  \bibinfo{pages}{253–--280}.
%Type = Article
\bibitem[{Varghese et~al.(2007b)Varghese, Frankel and
  Fischer}]{varghese_frankel_fischer_2007b}
\bibinfo{author}{Varghese, S.S.}, \bibinfo{author}{Frankel, S.H.},
  \bibinfo{author}{Fischer, P.F.}, \bibinfo{year}{2007}b.
\newblock \bibinfo{title}{Direct numerical simulation of stenotic flows. {Part}
  2. {Pulsatile} flow}.
\newblock \bibinfo{journal}{Journal of Fluid Mechanics} \bibinfo{volume}{582},
  \bibinfo{pages}{281–--318}.
%Type = Article
\bibitem[{van~de Vosse and
  Stergiopulos(2011)}]{van_de_Vosse_Stergiopoulos_2011}
\bibinfo{author}{van~de Vosse, F.N.}, \bibinfo{author}{Stergiopulos, N.},
  \bibinfo{year}{2011}.
\newblock \bibinfo{title}{Pulse wave propagation in the arterial tree}.
\newblock \bibinfo{journal}{Annual Review of Fluid Mechanics}
  \bibinfo{volume}{43}, \bibinfo{pages}{467--499}.
%Type = Article
\bibitem[{Young and Tsai(1973a)}]{Young_Tsai_1973a}
\bibinfo{author}{Young, D.}, \bibinfo{author}{Tsai, F.}, \bibinfo{year}{1973}a.
\newblock \bibinfo{title}{Flow characteristics in models of arterial stenoses
  — {I.} {Steady} flow}.
\newblock \bibinfo{journal}{Journal of Biomechanics} \bibinfo{volume}{6},
  \bibinfo{pages}{395--410}.
%Type = Article
\bibitem[{Young and Tsai(1973b)}]{Young_Tsai_1973b}
\bibinfo{author}{Young, D.}, \bibinfo{author}{Tsai, F.}, \bibinfo{year}{1973}b.
\newblock \bibinfo{title}{Flow characteristics in models of arterial stenoses
  — {II.} {Unsteady} flow}.
\newblock \bibinfo{journal}{Journal of Biomechanics} \bibinfo{volume}{6},
  \bibinfo{pages}{547--559}.

\end{thebibliography}



%\begin{thebibliography}{00}



%\end{thebibliography}
\end{document}